%% file: ANA-MUON-2024-01-PAPER.tex
\documentclass[cernpreprint, atlasdraft=false, UKenglish, texmf, orcidlogo]{atlasdoc}
\usepackage{atlaspackage}
\usepackage{atlasbiblatex}

\usepackage{atlasphysics}
\graphicspath{{logos/}{figures/}}

\usepackage{ANA-MUON-2024-01-PAPER-defs}
\AtlasTitle{Development and performance of a transformer-based lepton isolation algorithm with the ATLAS experiment}

\AtlasAbstract{%
The Prompt Lepton Isolation Tagger (PLIT) is a transformer-based
algorithm designed to distinguish prompt leptons (electrons and muons)
produced in the primary decays of electroweak gauge bosons from non-prompt
leptons produced in hadron decays and from misidentified leptons. It is primarily used in measurements targeting multilepton final
states with important backgrounds from non-prompt and misidentified
leptons. PLIT exploits information from charged-particle tracks associated
with the lepton candidate, together with lepton properties,
using a transformer neural network to identify correlations between
the tracks and improve the discrimination. Its performance is evaluated using simulated top-quark pair
production events and compared with that of previously used lepton isolation algorithms,
demonstrating improved prompt lepton efficiency while maintaining high rejection of non-prompt leptons.
This article presents the development, performance, and calibration
of PLIT for electrons and muons in proton--proton collisions at
a centre-of-mass energy of $\sqrt{s}=13.6$~TeV recorded by
the ATLAS experiment at the Large Hadron Collider.
Operating points are defined for both electrons and muons,
and their efficiencies are calibrated using $\ensuremath{Z\to l^+l^-}$
events in Run~3 data, providing data-to-simulation correction
factors for physics analyses.
}

\AtlasRefCode{MUON-2024-01}

\PreprintIdNumber{CERN-EP-2026-292}

\AtlasJournal{EPJC}
\AtlasCoverEgroupAnalysisTeam{atlas-muon-2024-01-analysis-team@cern.ch}

\hypersetup{pdftitle={ATLAS document},pdfauthor={The ATLAS Collaboration}}

\begin{document}

\maketitle

\tableofcontents

\section{Introduction}
%

\begin{figure}
\centering
\includegraphics[width=0.7\textwidth]{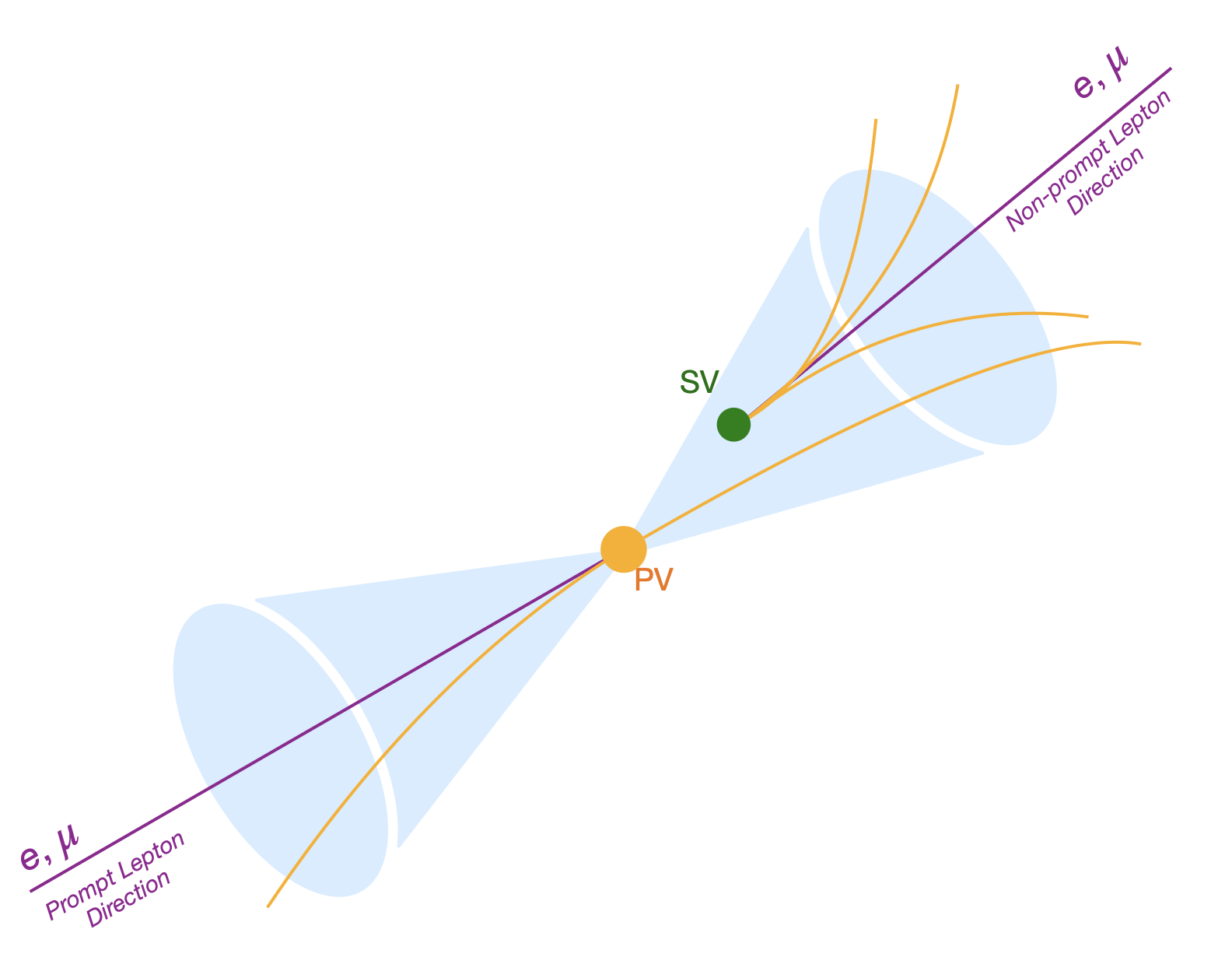}
\caption{Illustration of prompt and non-prompt leptons. Prompt leptons originate from the primary vertex (PV) and are typically surrounded by little nearby charged-particle activity. Non-prompt leptons predominantly originate from displaced secondary vertices (SV) from heavy-flavour hadron decays, and are generally accompanied by additional charged-particle tracks. The PLIT algorithm exploits conventional isolation observables together with the properties of tracks (shown as yellow lines) within a cone around the lepton candidate.}
\label{pnp}
\end{figure}

Many measurements at the LHC~\cite{Evans:2008zzb} rely on the identification of leptons (muons and electrons)
produced at the interaction point in the decay of electroweak gauge bosons ($W$ and $Z$).
Such leptons are prompt and are typically isolated from surrounding particle activity.
At the same time, LHC collisions produce a very large amount of leptons from heavy-flavour
($b$ and $c$) hadron decays, which constitute a background that needs to be rejected in many physics measurements. Such leptons are produced at a secondary vertex
displaced from the primary one, and are typically accompanied by nearby particles produced either in the same hadron
decay or in the same fragmentation process: they are non-prompt and non-isolated.
The efficient separation between these two sources of leptons, illustrated in Figure~\ref{pnp}, is particularly important for measurements targeting final states with two same-charge leptons or three or more leptons, commonly referred to as multilepton measurements, where other background processes are strongly suppressed by the lepton multiplicity and charge requirements.

Both the promptness and isolation properties can be used to achieve a good separation.
Conventional isolation requirements use variables based on the scalar sum of the transverse momentum of charged-particle tracks
and the transverse energy deposited in the calorimeter within a cone around the lepton candidate~\cite{MUON-2018-03, EGAM-2021-01}. The values of these variables,
commonly referred to as isolation observables, are required to remain below predefined thresholds.
Promptness is usually enforced by requiring the lepton to be compatible with originating from the primary vertex.
Multivariate isolation algorithms extend these approaches by combining conventional isolation observables with
lifetime information of the lepton and nearby charged particles to reject candidates produced in secondary vertices.

In most ATLAS measurements, muons and electrons from $\tau$ leptonic decays are not distinguished from those produced directly
in $W$ and $Z$ boson decays even though $\tau$-leptons have a significant lifetime: their contribution is typically small but
adds to the signal acceptance.
Therefore, throughout this article, ``prompt leptons'' refer to muons and electrons produced either directly in $W$ and $Z$ boson
decays at the primary vertex or in subsequent leptonic decays of $\tau$-leptons originating from these bosons.
In the measurements for which PLIT is intended,
light hadron decays, misidentified lepton candidates, and photon conversions misidentified as electrons,
constitute subdominant sources of background alongside the dominant contribution from heavy-flavour hadron decays;
these contributions are collectively referred to as ``non-prompt leptons''.

In the ATLAS measurement of the $t\bar{t}H$ process in multilepton final states using LHC Run~1 data~\cite{HIGG-2013-26}, the uncertainty on the non-prompt lepton background was the single largest systematic uncertainty. This motivated the development of the first multivariate lepton-isolation algorithms to further suppress this background. During Run~2, a boosted decision tree (BDT), referred to in its latest implementation as Prompt Lepton Improved Veto (PLIV), combined conventional tracking- and calorimeter-based isolation variables with jet flavour-tagging information associated with the jet closest to the lepton. The algorithm was successfully employed in several ATLAS measurements, including $t\bar{t}H$ multilepton measurements~\cite{HIGG-2018-13,HIGP-2024-08}, the $t\bar{t}W$ cross-section measurement~\cite{TOPQ-2019-30}, and the observation of four-top-quark production~\cite{TOPQ-2021-08}.
Given the large amount of non-prompt lepton background at low transverse momentum,
PLIV operated at low prompt lepton efficiency (below 50\%) in that region of phase space. Improving the prompt lepton efficiency while maintaining the same level of non-prompt lepton rejection would significantly increase the signal acceptance in multilepton measurements.

This paper presents the Prompt Lepton Isolation Tagger (PLIT), developed using a
transformer neural network architecture~\cite{vaswaniattentionneed} based on
the GN2 jet flavour-tagging algorithm developed by ATLAS~\cite{FTAG-2023-05}.
PLIT is designed to distinguish prompt leptons from non-prompt
leptons produced predominantly in semileptonic decays of heavy-flavour hadrons.
The training is performed using simulated $t\bar{t}$ events, with substantially
larger simulated samples available than for the training of the previous PLIV
algorithm.

PLIT combines conventional isolation observables with detailed
information from charged-particle tracks reconstructed within a cone around the lepton
candidate, using the same underlying track representation as the GN2 algorithm.
As a consequence, PLIT does not depend on jet reconstruction or jet
flavour-tagging tools.

The scope of this paper includes the development and performance of PLIT for
muons and electrons, the definition of the algorithm operating points (OPs),
and the derivation of the corresponding data-to-simulation efficiency
correction factors from proton--proton collision data. Separate PLIT models were trained for LHC Run~2 at $\sqrt{s}=13$~\TeV
and LHC Run~3 at $\sqrt{s}=13.6$~\TeV using the same architecture and
methodology, with the training adapted to the respective detector and
data-taking conditions. The focus of this paper is the Run~3
implementation of PLIT.

The paper is structured as follows.
The ATLAS detector is presented in Section~\ref{sec:detector}.
The reconstruction and selection of physics objects used in this study are described in Section~\ref{sec:objects}.
The recorded data and Monte Carlo (MC) simulated event samples used for the training, evaluation, and calibration of the algorithm are described in Section~\ref{sec:dataMC}.
The PLIT algorithm architecture, training procedure, and optimisation strategy are presented in Section~\ref{sec:plit}.
The performance of the algorithm in simulated events, including the definition and evaluation of the OPs, is described in Section~\ref{sec:plit_perf}.
The calibration of the algorithm and the derivation of data-to-simulation efficiency correction factors are presented in Section~\ref{sec:calibration}.
Finally, Section~\ref{sec:conclusion} summarises the results.


%
%
\section{ATLAS detector}
\label{sec:detector}
%

%
\newcommand{\AtlasCoordFootnote}{%
ATLAS uses a right-handed coordinate system with its origin at the nominal interaction point (IP)
in the centre of the detector and the \(z\)-axis along the beam pipe.
The \(x\)-axis points from the IP to the centre of the LHC ring,
and the \(y\)-axis points upwards.
Polar coordinates \((r,\phi)\) are used in the transverse plane,
\(\phi\) being the azimuthal angle around the \(z\)-axis.
The pseudorapidity is defined in terms of the polar angle \(\theta\) as \(\eta = -\ln \tan(\theta/2)\) and is equal to the rapidity
$ y = \frac{1}{2} \ln \left( \frac{E + p_z}{E - p_z} \right) $ in the relativistic limit.
Angular distance is measured in units of \(\Delta R \equiv \sqrt{(\Delta \eta)^{2} + (\Delta\phi)^{2}}\).}

The ATLAS detector~\cite{PERF-2007-01,GENR-2019-02} at the LHC covers nearly the entire solid angle around the collision point.
It consists of an inner tracking detector surrounded by a thin superconducting solenoid, electromagnetic and hadronic calorimeters,
and a muon spectrometer incorporating three large superconducting air-core toroidal magnets.

The inner-detector system (ID) is immersed in a \qty{2}{\tesla} axial magnetic field
and provides charged-particle tracking in the range \(|\eta| < 2.5\)\footnote{\AtlasCoordFootnote}.
The high-granularity silicon pixel detector covers the vertex region and typically provides four measurements per track~\cite{ATLAS-TDR-19,PIX-2018-001}. The IBL is the innermost pixel layer, followed by the B-layer. Beyond the pixel detector, the SemiConductor Tracker (SCT) typically provides eight measurements per track.
These silicon detectors are complemented by the transition radiation tracker (TRT),
which enables radially extended track reconstruction up to \(|\eta| = 2.0\).
The TRT also provides electron identification information
based on the fraction of hits (typically 30 in total) above a higher energy-deposit threshold corresponding to transition radiation.

The calorimeter system covers the pseudorapidity range \(|\eta| < 4.9\).
Within the region \(|\eta|< 3.2\), electromagnetic calorimetry is provided by barrel and
endcap high-granularity lead/liquid-argon (LAr) calorimeters,
with an additional thin LAr presampler covering \(|\eta| < 1.8\)
to correct for energy loss in material upstream of the calorimeters.
Hadronic calorimetry is provided by the steel/scintillator-tile calorimeter,
segmented into three barrel structures within \(|\eta| < 1.7\), and two copper/LAr hadronic endcap calorimeters.
The solid angle coverage is completed with forward copper/LAr and tungsten/LAr calorimeter modules
optimised for electromagnetic and hadronic energy measurements respectively.

The muon spectrometer (MS) comprises separate trigger and
high-precision tracking chambers measuring the deflection of muons in a magnetic field generated by the superconducting air-core toroidal magnets.
The field integral of the toroids ranges between \num{2.0} and \qty{6.0}{\tesla\metre}
across most of the detector.
Three layers of precision chambers, each consisting of layers of monitored drift tubes (MDT), cover the region \(|\eta| < 2.7\),
except for the innermost layer in the endcap region, where the background is highest: in Run~2 the MDTs were complemented by cathode-strip chambers in the forward region,
while in Run~3 small-strip Thin-Gap Chambers (sTGC) and Micromegas (MM) chambers provide precision tracking in the region \(1.3 < |\eta| < 2.7\)~\cite{GENR-2019-02}.
The muon trigger system covers the range \(|\eta| < 2.4\) with resistive-plate chambers in the barrel, while in the endcap it is based on Thin-Gap Chambers (TGC),
with sTGC and MM chambers providing the trigger in the innermost endcap layer.

The luminosity is measured mainly by the LUCID--2~\cite{LUCID2} detector that records Cherenkov light produced in the quartz windows of photomultipliers located close to the beampipe.

Events are selected by the first-level trigger system implemented in custom hardware,
followed by selections made by algorithms implemented in software in the high-level trigger~\cite{TRIG-2016-01,TRIG-2022-01}.
The first-level trigger accepts events from the \qty{40}{\MHz} bunch crossings at a rate below \qty{100}{\kHz},
which the high-level trigger further reduces to record complete events to disk at about \qty{1.25}{\kHz} in Run~2 and \qty{3}{\kHz} in Run~3.

A software suite~\cite{SOFT-2022-02} is used in data simulation, in the reconstruction
and analysis of real and simulated data, in detector operations, and in the trigger and data acquisition
systems of the experiment.


%
\section{Object reconstruction and selection}
\label{sec:objects}
%

Interaction vertices from $pp$ collisions are reconstructed from at least two tracks with $\pt > 500$~\MeV that are consistent with originating from the beam collision region in the $x$--$y$ plane. If more than one primary-vertex candidate is reconstructed in an event, the hard-scatter primary vertex is defined as the candidate for which the associated tracks have the largest summed squared transverse momentum. More details about track and vertex reconstruction in ATLAS can be found in~ Ref.~\cite{IDTR-2021-01}.

Muon candidates are reconstructed by performing a combined fit of the track information from the ID and MS~\cite{MUON-2018-03}.
Muons are required to satisfy the \textsc{Medium} identification OP and satisfy $\pt > 10~\gev$ and $|\eta| < 2.5$, as well as a transverse impact parameter significance of $|d_{0}|/\sigma(d_{0}) < 3$ and a longitudinal impact parameter relative to the primary vertex of $|z_{0}\sin\theta| < 0.5$~mm.

Electron candidates are reconstructed from energy clusters in the electromagnetic
calorimeter matched to a track (or tracks) in the ID~\cite{EGAM-2021-01}.
Either the \textsc{Loose} or \textsc{Medium} electron identification
OP~\cite{EGAM-2021-01} is used, as specified in the relevant sections.
These identification OPs are based on a likelihood discriminant employing
calorimeter and tracking variables that provide separation between electrons
and jets, and between electrons and photons.
Electrons are required to satisfy $\pt>10$~\GeV and
$|\eta|<2.47$, excluding the transition region between the
endcap and barrel calorimeters ($1.37<|\eta|<1.52$), as well
as $|d_{0}|/\sigma(d_{0}) < 5$ and $|z_{0}\sin\theta| < 0.5$~mm with regard to the
event primary vertex.


%
\section{Data and simulated event samples}
\label{sec:dataMC}
%

The PLIT algorithm is trained and its performance is evaluated using simulated
top-quark pair ($\ttbar$) production events. Besides being a major background in many
multilepton analyses, $t\bar{t}$ events naturally provide both prompt and
non-prompt leptons. All $t\bar{t}$
decay modes are considered: leptonic $W$ boson decays provide prompt leptons,
while $b$-jets and hadronic $W$ boson decays provide a large sample of
non-prompt leptons, predominantly from heavy-flavour hadron decays. The calibration of the prompt lepton PLIT efficiency is derived using simulated and real $\Zmm$ and $\Zee$ events.

The production of $\ttbar$ events in simulation was modelled using the
\POWHEGBOX[v2] generator at next-to-leading-order (NLO) in
QCD~\cite{Frixione:2007nw,Nason:2004rx,Frixione:2007vw},
with the NNPDF3.0 NLO~\cite{Ball:2014uwa} parton distribution function (PDF) set.
The \hdamp parameter\footnote{The \hdamp parameter is a resummation
damping factor and one of the parameters that controls the matching
of \POWHEG matrix elements to the parton shower and thus effectively
regulates the high-\pT radiation against which the hard-process system
recoils.}
was set to 1.5 times the top-quark mass ($m_{\mathrm{top}}$).
The functional form of the renormalisation and factorisation scales was set
to the default scale of $\sqrt{m_{\mathrm{top}}^{2}+p_{\mathrm{T},t}^{2}}$,
where the top-quark transverse momentum $p_{\mathrm{T},t}$ was evaluated before radiation.
The events were interfaced with \PYTHIA[8.245]~\cite{Sjostrand:2014zea}
using the A14~\cite{ATL-PHYS-PUB-2014-021} tuned parameters and the
NNPDF2.3 LO PDF set~\cite{Ball:2012cx}, while heavy-flavour hadron
decays were modelled using \textsc{EvtGen}~1.2.0~\cite{Lange:2001uf}.

Samples of $\Zmm$ and $\Zee$ events were simulated at NLO accuracy in perturbative QCD using \POWHEGBOX[v1]~\cite{Nason:2004rx,Frixione:2007vw,Alioli:2010xd,Alioli:2008gx}, with the \CT[10nlo] PDF set~\cite{Lai:2010vv}, interfaced to \PYTHIA[8.245]~\cite{Sjostrand:2014zea}, with the \CTEQ[6L1] PDF set~\cite{Pumplin:2002vw} and the \AZNLO~\cite{STDM-2012-23} tuned generator parameters. \PHOTOSpp{} was used for simulating the QED final-state radiation. Alternative samples of $\Zmm$ and $\Zee$ events were simulated with \SHERPA[2.2.11]~\cite{Bothmann:2019yzt,PMGR-2021-01} using NLO matrix element for up to two partons and leading-order matrix element for up to five partons, calculated with \textsc{Comix}~\cite{Gleisberg:2008fv} and \OPENLOOPS~\cite{Buccioni:2019sur,Cascioli:2011va,Denner:2016kdg}. For these samples, the \SHERPA parton shower~\cite{Schumann:2007mg} was used, with the set of tuned parameters developed by the \SHERPA authors. The \NNPDF[3.0nnlo] PDF set~\cite{Ball:2014uwa} was used.

All simulated samples were processed through a detailed detector
simulation based on \textsc{Geant4}~\cite{Agostinelli:2002hh}.
The effect of multiple \pp interactions in the same and neighbouring
bunch crossings (pile-up) was included in the simulation~\cite{SIMU-2020-01}.
The simulated events were reweighted to match the distribution of the
average number of interactions per bunch crossing, $\langle\mu\rangle$,
observed in data, which is typically around 60 for the Run~3 dataset.
Simulations were produced separately at centre-of-mass energies of
13~\TeV and 13.6~\TeV, corresponding to LHC Run~2 and Run~3, respectively,
and were reconstructed using detector and pile-up conditions
representative of the corresponding data-taking periods.

The data used to derive data-to-simulation correction factors for the
PLIT efficiency for muons and electrons were collected in proton--proton
collisions at a centre-of-mass energy of $\sqrt{s}=13.6$~\TeV during
2022--2024. Only data recorded during periods with stable LHC beams
and for which all relevant ATLAS detector systems were operational
were considered, following standard ATLAS data-quality
requirements~\cite{SOFT-2022-02,DAPR-2018-01}.
Events were selected using the lowest unprescaled single-muon and
single-electron triggers~\cite{TRIG-2018-05,TRIG-2018-01,TRIG-2022-01}.
The resulting dataset corresponds to a total integrated luminosity of
{$166 \pm 3$}~\ifb for muons and {$165 \pm 3$}~\ifb for electrons~\cite{ATL-DAPR-PUB-2023-001,ATL-DAPR-PUB-2024-001,ATL-DAPR-PUB-2025-001}.


%
\section{The PLIT transformer neural network: architecture and training}
\label{sec:plit}
%

The PLIT architecture is based on the GN2 transformer architecture developed in ATLAS for heavy-flavour jet tagging~\cite{FTAG-2023-05}, adapted to the problem of lepton isolation. The Salt framework~\cite{salt2025} is used to train the transformer.

The algorithm processes the kinematic and detector-related properties
of the lepton together with information from tracks reconstructed in
its vicinity: for each lepton candidate, up to 40 tracks within a cone
of $\Delta R < 0.4$ around the lepton are associated with the
lepton. In practice, this limit has no impact, since fewer than 40 tracks
are always found. The same track requirements as those used for the GN1 flavour-tagging algorithm~\cite{ATL-PHYS-PUB-2022-027} are applied before the association: they are required to satisfy $\pt > 500$~\MeV and $|\eta| < 2.5$, $|d_{0}| < 3.5$~mm and $|z_{0}\sin\theta| < 5$~mm with regard to the event primary vertex, and pass certain quality criteria in terms of hit multiplicity and quality in the Pixel and SCT silicon detectors, as shown in Table~\ref{table:track_quality}.
Split hits are hits shared by multiple tracks that are classified as split
by the cluster-splitting neural network~\cite{PERF-2015-08}.
Shared hits are hits shared by multiple tracks that are not classified
as split by this network. The total number of shared hits is evaluated by giving a weight of 1 to shared hits in pixel layers,
and a weight of 0.5 to shared hits in the SCT. Hits expected in non-functioning detector modules are included
in the total silicon hit count used for track selection,
so that tracks are not penalised for passing through these modules.
A hole is defined as a missing hit in a detector layer where
a hit is expected, between two layers with recorded hits on the track.

\begin{table}[!htbp]
\caption{Quality selections applied to tracks associated with the leptons.}
\label{table:track_quality}
\small
\centering
\begin{tabular}{lp{0.12\linewidth}}

\textbf{Variable} & \textbf{Selection} \\
\hline
$p_{\textrm{T}}$       & $> 500$~\MeV  \\
$|\eta|$            & $<2.5$       \\
$|d_{0}|$            & $< 3.5$~mm   \\
$|z_{0}\sin\theta|$  & $<5$~mm      \\
Silicon hits (incl. dead modules) & $\ge 8$       \\
Shared silicon hits & $< 2$       \\
Silicon holes       & $< 3$       \\
Pixel holes         & $< 2$       \\
\end{tabular}
\end{table}

The input features used for the training are summarised in
Table~\ref{table:track_inputs}. The inputs are divided into two
categories: variables describing the reconstructed lepton and
variables describing the associated tracks. The lepton-level inputs include kinematic properties together with the conventional track- and calorimeter-based isolation observables, while the track-level inputs comprise reconstructed track parameters, relative track--lepton kinematics, and variables related to track quality.

\begin{table}[!htbp]
\caption{Input features to the PLIT algorithm for the lepton and up to 40 tracks within $\Delta R < 0.4$ of the lepton.}
\label{table:track_inputs}
\small
\centering
\begin{tabular}{p{0.2\linewidth}p{0.6\linewidth}}

\textbf{Lepton-level Input} & \textbf{Description} \\
\hline
$p_{\textrm{T}}$ & Transverse momentum \\
$\eta$ & Pseudorapidity \\
$\phi$ & Azimuthal angle \\
$p_{\textrm{T}}^{\textrm{varcone30}} / p_{\textrm{T}}$ & Scalar sum of the $p_{\textrm{T}}^{\text{track}}$ of the inner detector tracks associated with the primary vertex in a cone of size $\Delta R < \min(10\, \mathrm{GeV}/ \pt, 0.3)$ around the lepton, excluding the lepton track itself, divided by the lepton $p_{\textrm{T}}$ \\
$E_{\textrm{T}}^{\textrm{topocone30}} / p_{\textrm{T}}$ & Sum of the transverse energy of topological clusters in a cone of size $\Delta R = 0.3$ around the lepton, after subtracting the contribution from the energy deposit of the lepton itself and correcting for pile-up effects, divided by the lepton $p_{\textrm{T}}$~\cite{MUON-2018-03, EGAM-2021-01}\\
caloClusterERel (muons only)         & Sum of the energies of cells along the muon path divided by the muon's expected energy loss in the calorimeter \\
caloClusterSumEtRel (electrons only) & Sum of the transverse energies of positive-energy topological clusters within $\Delta R < 0.15$ around the electron divided by the electron $p_{\textrm{T}}$\\
\\
\textbf{Track-level Input} & \textbf{Description} \\
\hline
$q^{\text{track}} / p^{\text{track}}$ & Track charge divided by momentum \\
$d_0$ & Transverse impact parameter measured with respect to the primary vertex \\
$z_0 \sin \theta$ & Longitudinal impact parameter times sine of the polar angle measured with respect to the primary vertex \\
$d_0/\sigma(d_0)$ & Significance of the transverse impact parameter \\
$z_0 \sin \theta/\sigma(z_0 \sin \theta)$ & Significance of the longitudinal impact parameter times sine of the polar angle \\
$\Delta \eta(\text{lepton}, \textrm{track})$ & Difference in $\eta$ between the lepton and the track \\
$\Delta \phi(\text{lepton}, \textrm{track})$ & Difference in $\phi$ between the lepton and the track \\
nPixHits & Number of pixel hits \\
nIBLHits & Number of IBL hits \\
nBLHits & Number of B-layer hits \\
nIBLShared & Number of shared IBL hits \\
nIBLSplit & Number of split IBL hits \\
nPixShared & Number of shared pixel hits \\
nPixSplit & Number of split pixel hits \\
nSCTHits & Number of SCT hits \\
nSCTShared & Number of shared SCT hits \\
lepton\_track & Indicates whether the track was used in the reconstruction of the lepton \\
\end{tabular}
\end{table}

The primary objective of PLIT is the classification of leptons into prompt and non-prompt categories. The network outputs the prompt lepton probability, denoted by $p_{\mathrm{prompt}}$, from which the PLIT discriminant is constructed as

\begin{equation}
D_{\mathrm{PLIT}} = \log \frac{p_{\mathrm{prompt}}}{1 - p_{\mathrm{prompt}}}.
\end{equation}

$D_{\mathrm{PLIT}}$ is more practical to use than $p_{\mathrm{prompt}}$ because the algorithm typically operates at values of $p_{\mathrm{prompt}}$ that are very close to 1.
In addition to the primary classification objective, two auxiliary
objectives are employed during training.
The first classifies each track according to its production origin (e.g. prompt, heavy-flavour decay, etc.),
while the second identifies groups of tracks originating from a common
production vertex, enabling the network to learn vertexing information
implicitly. These auxiliary objectives improve the representation
learning capabilities of the network. Their outputs are not used in the construction of the PLIT discriminant,
but can be inspected separately to enhance the interpretability of the network predictions.

A schematic overview of the PLIT architecture is shown in
Figure~\ref{PLITarchitecture}. For each associated track, the
lepton-level and track-level input features are first concatenated to
form a combined input vector. These combined input vectors are then
passed through a fully connected embedding network consisting of a
single hidden layer with a 256-dimensional output.
This embedding step transforms the input features into
a fixed-dimensional representation for each track, allowing the
information associated with different tracks to be processed together.
These initial track representations are then processed by a
transformer encoder composed of three stacked transformer layers with
eight attention heads. Through the self-attention mechanism,
the transformer allows the representation of each track to incorporate
information from other tracks associated with the same lepton. This
enables the network to account for relationships between tracks rather
than processing each track independently. The use of multiple attention
heads allows different patterns of relationships between tracks to be
considered in parallel. The transformer operates on embeddings of
dimension 256 using the pre-LayerNorm configuration, in which
normalisation is applied before the attention and feed-forward
operations. The resulting track representations are combined into
a global lepton representation using attention pooling, which is then
used by the primary classification network to predict the prompt
and non-prompt lepton probabilities.

Additionally, the auxiliary objectives exploit the individual track
embeddings. The track-origin classification task predicts eight
track-origin categories~\cite{ATL-PHYS-PUB-2026-001}, distinguishing tracks originating from direct $b$-hadron decays,
$c$-hadrons produced in $b$-hadron decays, direct $c$-hadron decays, $\tau$-lepton decays,
the primary vertex, pile-up interactions, and tracks that do not correspond to genuine
charged-particle trajectories. Tracks that cannot be assigned to any of these categories are grouped into an
``other'' category, including tracks from light-hadron decays and interactions
in the detector. The vertexing task operates on
pairs of track embeddings. It predicts whether pairs of tracks
originate from a common vertex, providing information about the spatial
relationships between tracks. ReLU activations and cross-entropy loss
functions are used throughout the architecture. The final loss is
constructed as a weighted combination of the three task losses,
enabling simultaneous optimisation of the full network.

\begin{figure}[h]
\centering
\includegraphics[width=0.99\textwidth]{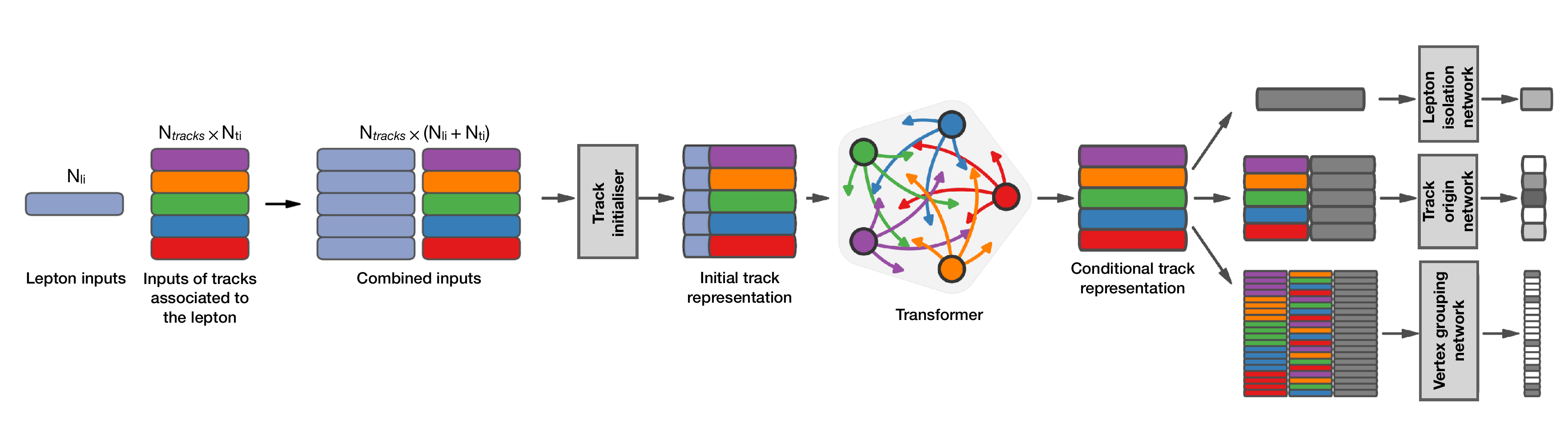}
\caption{Schematic overview of the PLIT architecture, based on the GN2 transformer architecture developed in ATLAS for heavy-flavour jet tagging~\cite{FTAG-2023-05}. $\mathrm{N_{li}}$ and $\mathrm{N_{ti}}$ correspond respectively to the number of lepton-level and track-level inputs. $\mathrm{N_{tracks}}$ is the number of tracks associated with a given lepton. Combined lepton and track features are processed by a transformer encoder to produce conditional track representations, which are subsequently used for prompt lepton classification together with auxiliary track-origin and vertex prediction tasks.}
\label{PLITarchitecture}
\end{figure}

Separate networks are trained on simulated $\ttbar$ events for muons, barrel electrons
($|\eta|<1.37$), and endcap electrons
($1.52<|\eta|<2.47$), using all the leptons satisfying the selection
described in Section~\ref{sec:objects}. Muons satisfying the \textsc{Medium}
identification OP and electrons satisfying the \textsc{Loose} identification OP
are used for training and performance evaluation, with the looser electron
selection providing increased training sample size. Samples are ten times larger than the ones
used previously to train PLIV; they contain 75 million muons and 68 million electrons for training, and 15 million muons and
10 million electrons for validation and testing, with a proportion of 80\% of prompt and 20\% of non-prompt leptons.

Figures~\ref{fig:bkg_comp_100_muon},~\ref{fig:bkg_comp_100_barrel} and~\ref{fig:bkg_comp_100_endcap} show the distribution of the PLIT discriminant in simulated $\ttbar$ events for muons, barrel electrons, and endcap electrons.
The prompt lepton contribution is compared with categories of non-prompt leptons originating from decays of $b$-hadrons, $c$-hadrons, light hadrons, misidentified lepton candidates, $\tau$-leptons decaying hadronically, and (in the case of electrons) photon conversions, showing the discriminating power of the algorithm.
Leptons from light-hadron decays and misidentified lepton candidates are grouped into a single category.
In this \ttbar sample, the dominant source of background comes from heavy-flavour ($b$- and $c$-) hadron decays.
In the case of electrons, light hadron decays, misidentified lepton candidates and conversions are also important;
PLIT is able to discriminate them as well, albeit with less discriminating power than the ones from heavy-flavour decays.

\begin{figure}[h!]
\centering
\includegraphics[width=0.8\textwidth]{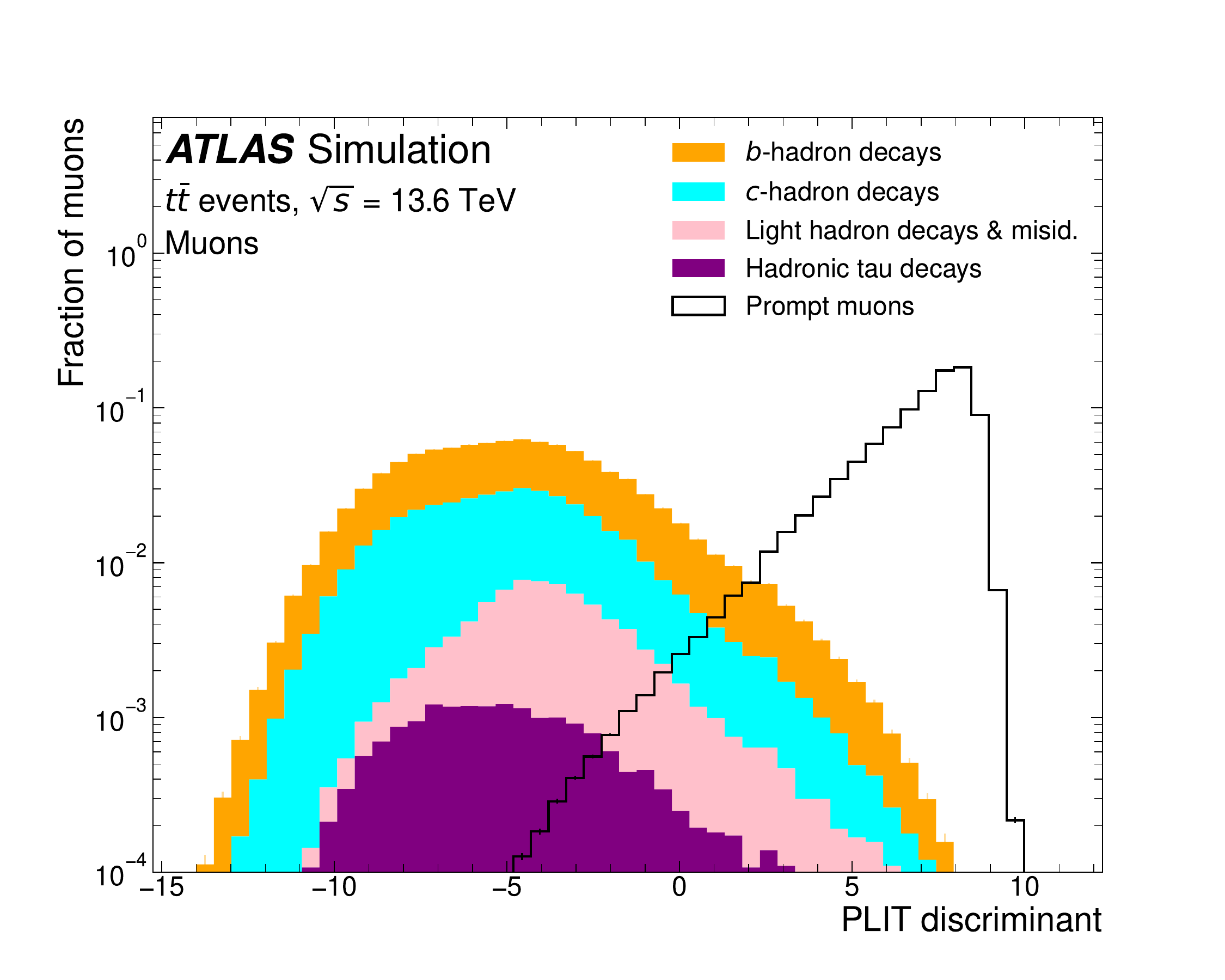}
\caption{PLIT discriminant in simulated $\ttbar$ events for non-prompt
(stacked histogram) and prompt (line) muons satisfying the \textsc{Medium}
identification OP.}
\label{fig:bkg_comp_100_muon}
\end{figure}

\begin{figure}[h!]
\centering
\includegraphics[width=0.8\textwidth]{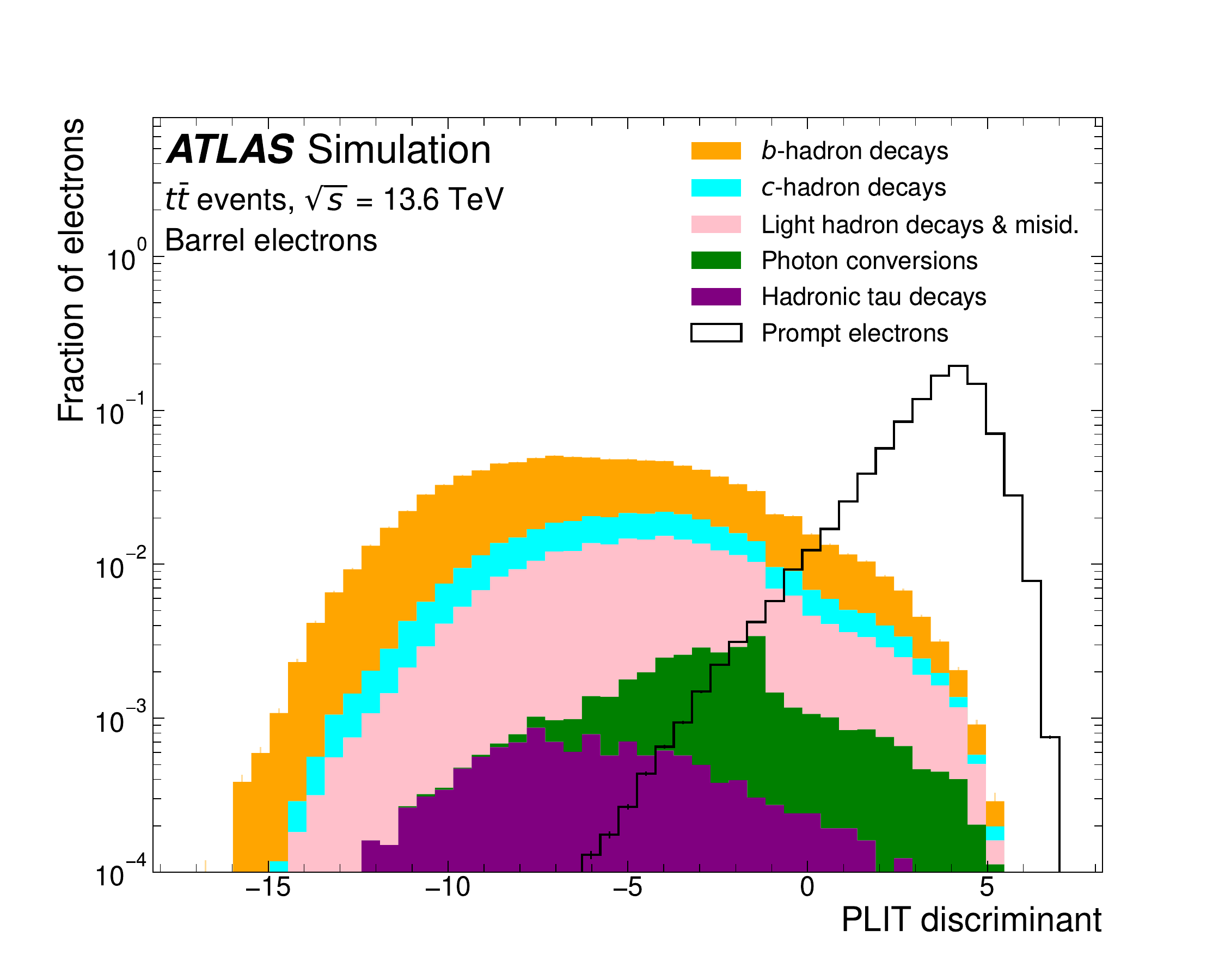}
\caption{PLIT discriminant in simulated $\ttbar$ events for non-prompt
(stacked histogram) and prompt (line) barrel electrons satisfying the \textsc{Loose}
identification OP.}
\label{fig:bkg_comp_100_barrel}
\end{figure}

\begin{figure}[h!]
\centering
\includegraphics[width=0.8\textwidth]{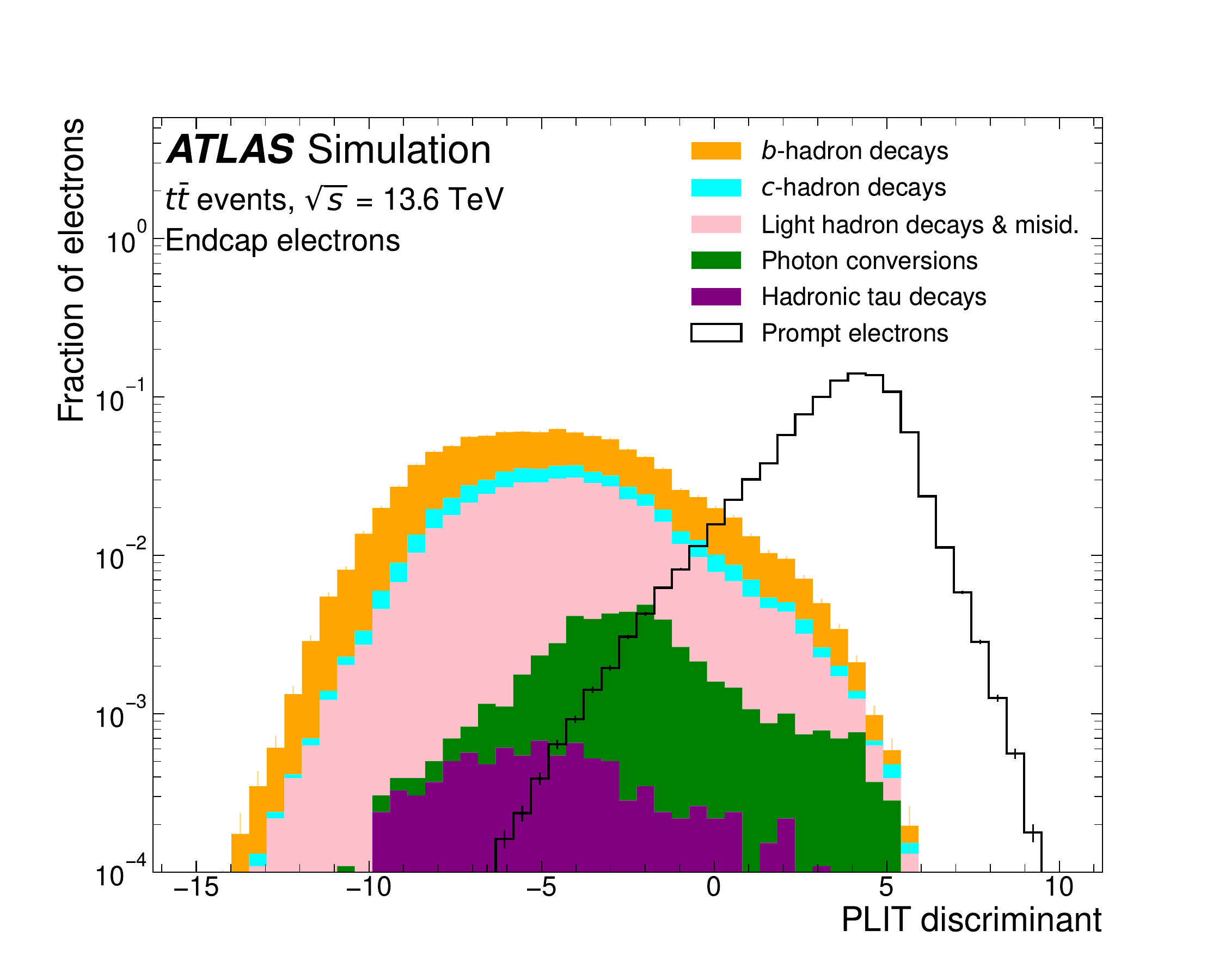}
\caption{PLIT discriminant in simulated $\ttbar$ events for non-prompt
(stacked histogram) and prompt (line) endcap electrons satisfying the \textsc{Loose}
identification OP.}
\label{fig:bkg_comp_100_endcap}
\end{figure}


%
\section{Performance of the PLIT algorithm}
\label{sec:plit_perf}
\subsection{ROC curve performance}

The PLIT performance is evaluated on the same simulated $\ttbar$ samples used for the training, albeit on a statistically independent sub-sample.
Figures~\ref{fig:muon_publiccc},~\ref{fig:barrel_publiccc}, and~\ref{fig:endcap_publiccc} show the prompt lepton efficiency vs non-prompt lepton rejection (ROC curve) for PLIV and PLIT, respectively for muons, barrel electrons, and endcap electrons.
The PLIV algorithm was not retrained for Run 3, so the Run~2 training is evaluated on Run~3 samples.
It was checked that the PLIV performance is only mildly degraded in Run~3 compared with Run~2, showing good resilience to the increased pile-up.
Since the performance varies significantly with $\pt$, it is shown separately for lepton $\pt$ below and above 20~\GeV. For comparison, the \loose and \tight OPs of the conventional lepton isolation combining track-based and calorimeter-based isolation selections~\cite{MUON-2018-03,EGAM-2021-01} are shown as well. Both PLIV and PLIT outperform significantly the standard isolation OPs across the entire $\pt$ spectrum.
PLIT performs better than PLIV in both $\pt$ ranges, across the entire efficiency range, especially where it matters most: for muons, at low $\pt$,
the PLIT rejection is between 1.5 and 2 times larger than the PLIV one for any given efficiency.
For electrons, the gain in rejection is a factor 2 or more in both $\pt$ ranges, and up to a factor 5 for endcap electrons at low efficiency.

\begin{figure}[h!]
\centering
\includegraphics[width=0.8\textwidth]{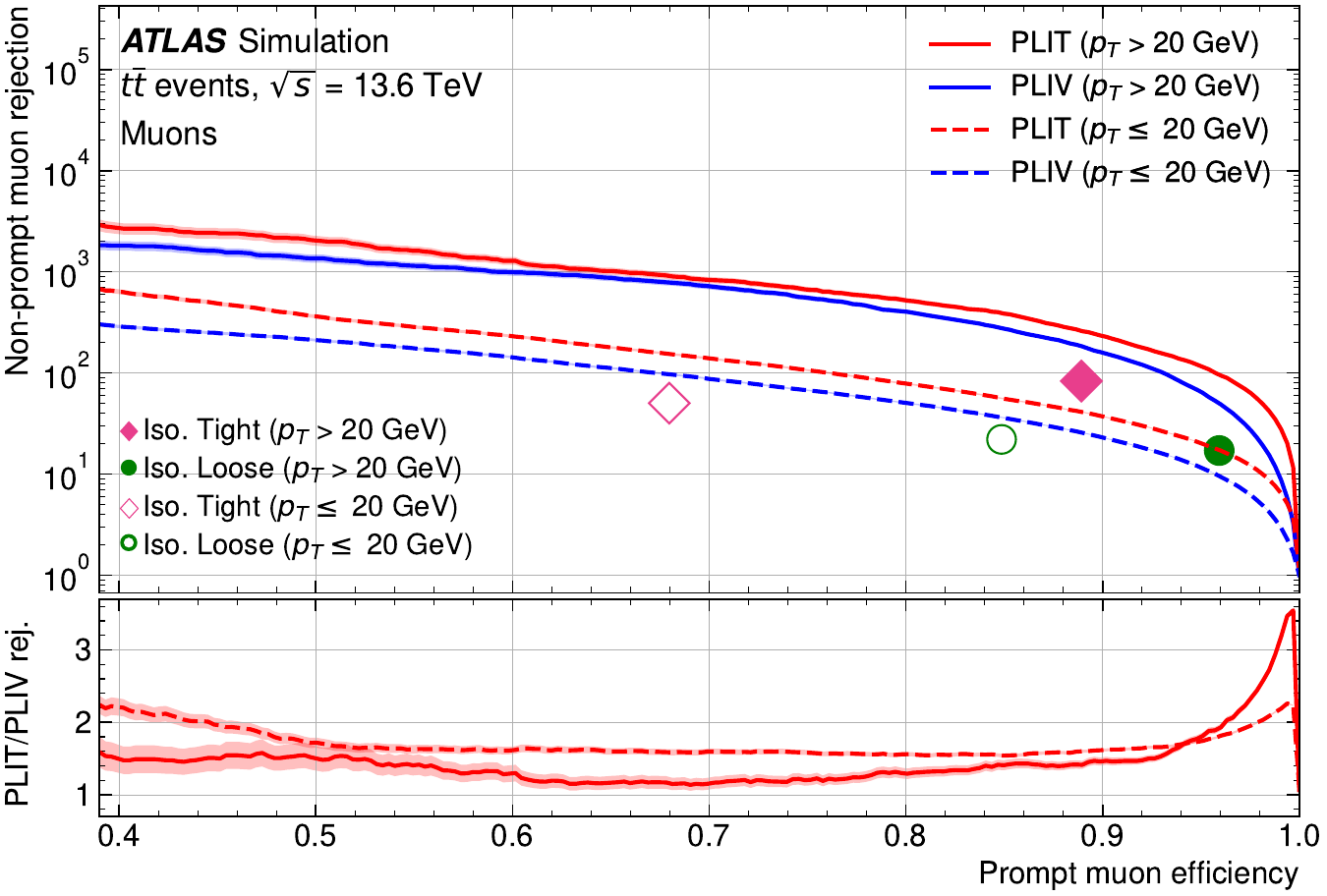}
\caption{PLIT and PLIV performance in $\ttbar$ Run~3
simulation for muons satisfying the \textsc{Medium} identification OP.
The non-prompt muon rejection is shown as a function of the prompt muon efficiency
separately for muon transverse momentum below (dashed lines) and above (solid lines) 20~\GeV.
The bottom panel shows the PLIT-to-PLIV rejection ratio.}
\label{fig:muon_publiccc}
\end{figure}

\begin{figure}[h!]
\centering
\includegraphics[width=0.8\textwidth]{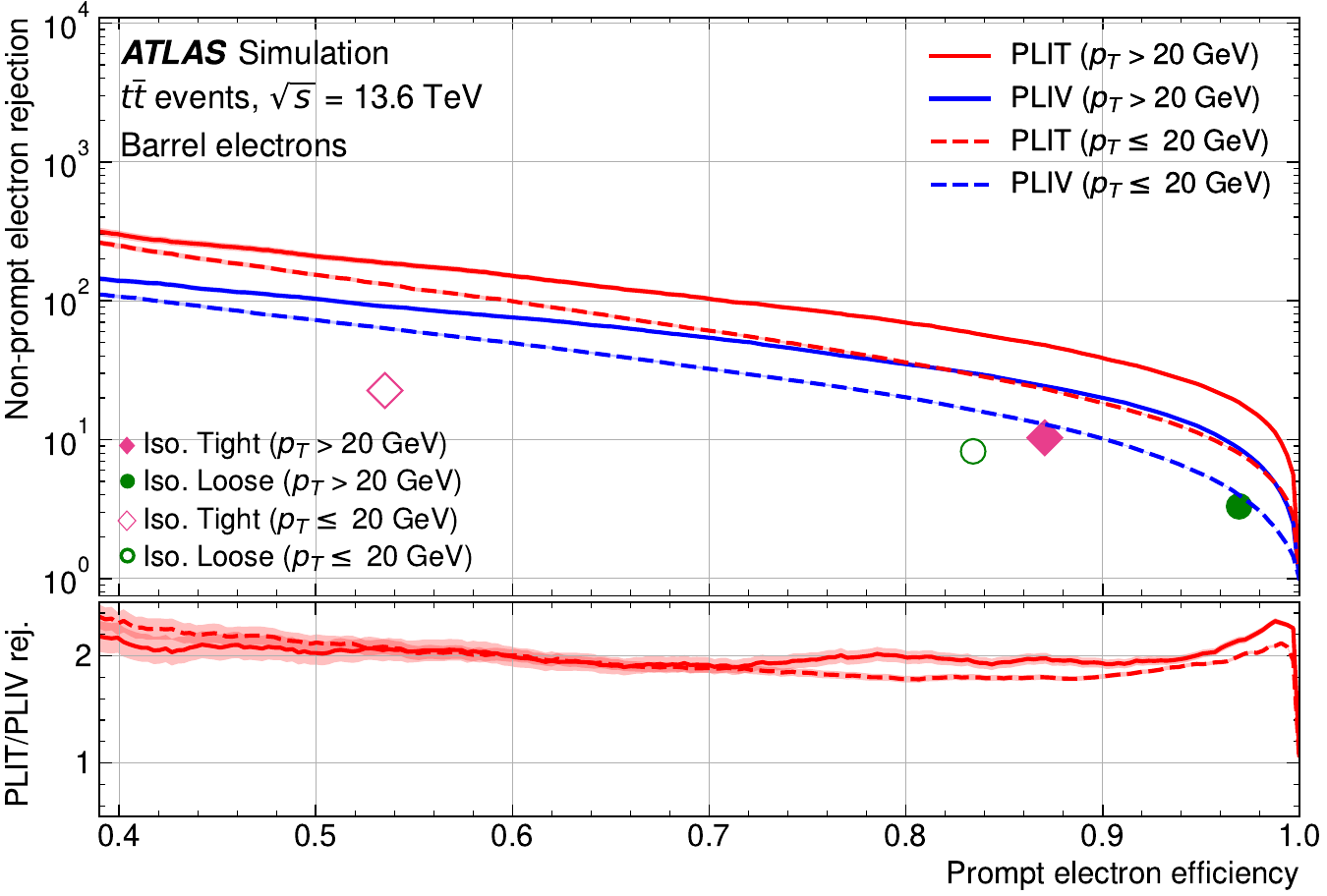}
\caption{PLIT and PLIV performance in $\ttbar$ Run~3
simulation for barrel electrons satisfying the \textsc{Medium} identification OP.
The non-prompt electron rejection is shown as a function of the prompt electron efficiency
separately for electron transverse momentum below (dashed lines) and above (solid lines) 20~\GeV.
The bottom panel shows the PLIT-to-PLIV rejection ratio.}
\label{fig:barrel_publiccc}
\end{figure}

\begin{figure}[h!]
\centering
\includegraphics[width=0.8\textwidth]{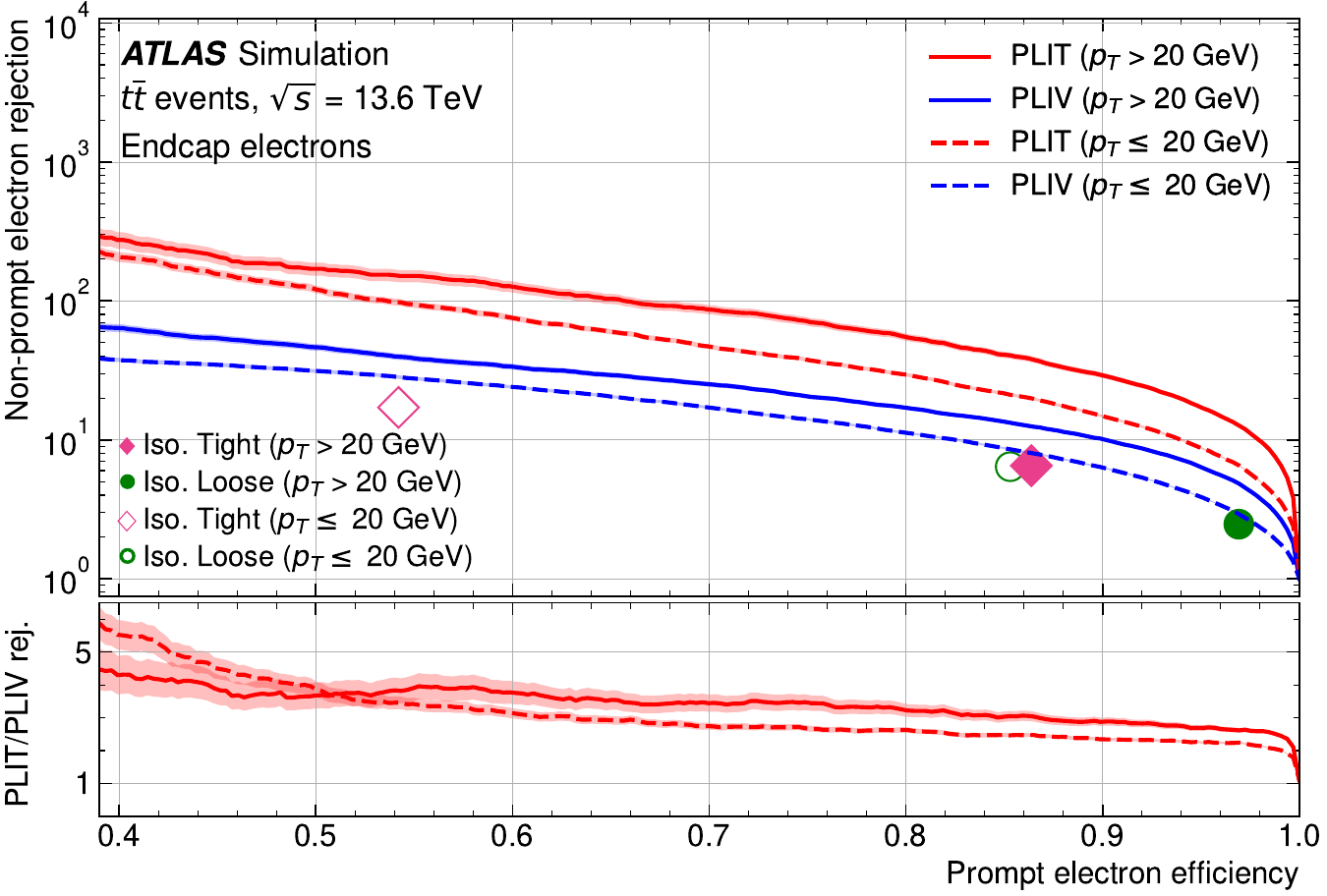}
\caption{PLIT and PLIV performance in $\ttbar$ Run~3
simulation for endcap electrons satisfying the \textsc{Medium} identification OP.
The non-prompt electron rejection is shown as a function of the prompt electron efficiency
separately for electron transverse momentum below (dashed lines) and above (solid lines) 20~\GeV.
The bottom panel shows the PLIT-to-PLIV rejection ratio.}
\label{fig:endcap_publiccc}
\end{figure}


\subsection{Definition and performance of the operating points}

The Tight and VeryTight OPs of the Run~2 PLIV BDT were originally defined to maximise the expected signal significance in the $t\bar{t}H$ and $t\bar{t}W$ multilepton analyses, respectively. Using Run~2 simulated events and the corresponding signal-region selections, the optimal PLIV threshold was determined as a function of the lepton transverse momentum, with separate optimisations performed for muons, barrel electrons, and endcap electrons. Since non-prompt leptons have a softer $\pt$ spectrum than the signal leptons, and the separation between prompt and non-prompt leptons is more challenging at low $\pt$, the resulting OPs become progressively tighter towards low $\pt$, leading to a lower prompt lepton efficiency in this region.

This optimisation is reused for PLIT by applying the same $\pt$-dependence to PLIT. For each PLIV OP (Tight and VeryTight), two PLIT OPs are defined: one matching the PLIV efficiency (leading to an improved rejection) and the other matching the PLIV rejection (leading to an improved efficiency). Prompt and non-prompt muons from simulated $\ttbar$ events, selected with the same criteria as described in Section~\ref{sec:detector}, are used both to define and evaluate the muon operating points.
For electrons, the PLIT operating points are defined by matching the prompt electron efficiency in simulated $\Zee$ events to that of the corresponding PLIV operating points, while the non-prompt rejection is defined using simulated $\ttbar$ events in which both $W$ bosons decay hadronically. The PLIT OPs for electrons are defined using the \textsc{Medium} electron identification OP. Since physics analyses may use different electron identification requirements, the dependence of the PLIT OP definition on the electron identification OP was studied and found to be small, not requiring a separate PLIT OP optimisation for each identification OP.

Figure~\ref{fig:wp_mu} shows the prompt muon efficiency (left) and non-prompt muon rejection (right) of the PLIT OPs in simulated Run~3 $\ttbar$ events, compared with the corresponding PLIV OPs, as a function of the muon transverse momentum (top), pseudorapidity (middle), and average number of interactions per bunch crossing (bottom).
By definition, the PLIV and PLIT efficiencies are identical for the matching-efficiency OP,
and their rejections are identical for the matching-rejection OP.
For the matching-rejection OP, the efficiency rises from 40\% (resp. 30\%) at $\pt=$10~\GeV
to more than 95\% (resp. 90\%) above 60~\GeV for the Tight (resp. VeryTight) OP. The steep efficiency turn-on curve
is a consequence of the tighter selection at low $\pt$. The non-prompt rejection reaches 350 (resp. 700)
for a prompt efficiency of 55\% (resp. 40\%) near $\pt \approx 20$~\GeV for the Tight (resp. VeryTight) OP.
Compared with PLIV, the relative efficiency gain is 10\% at low $\pt$ for the same rejection,
and the rejection improvement is between 20\% and 60\% as a function of $\pt$
for the same efficiency, with most of the gain achieved at higher $\eta$.
The dependence on the muon pseudorapidity provides additional insight into the behaviour of the globally defined OPs. While the prompt muon efficiency decreases towards larger $|\eta|$ values for both PLIT and PLIV, the non-prompt muon rejection achieved by PLIT increases with $|\eta|$, whereas the corresponding PLIV rejection decreases. As a function of the average number of interactions per bunch crossing, the prompt muon efficiency decreases only mildly owing to the increased detector activity at higher pile-up, while the rejection remains constant, showing good robustness against pile-up.

Similarly, Figures~\ref{fig:wp_elec_pt} and~\ref{fig:wp_elec_eta_mu} show the prompt electron efficiency (left) and the non-prompt electron rejection (right) for the PLIT and PLIV OPs, evaluated in simulated $\ttbar$ events, as a function of the electron transverse momentum (Figure~\ref{fig:wp_elec_pt}), pseudorapidity and the average number of interactions per bunch crossing (Figure~\ref{fig:wp_elec_eta_mu}).
The prompt electron efficiency is evaluated in simulated $\ttbar$ single-lepton events, while the non-prompt electron rejection is evaluated in simulated $\ttbar$ all-hadronic events. The improvement over PLIV is significant.
For the same efficiency, the rejection is improved by roughly a factor 2, and up to a factor 5 for endcap electrons at low $\pt$.
For the same rejection, the relative efficiency gain at low $\pt$ is 50\% in the barrel, and a factor 2 in the endcap.
As a function of pseudorapidity, the same behaviour as for muons is observed: the prompt electron efficiency exhibits only a mild dependence, with a decrease beyond the barrel--endcap transition region, while the PLIT rejection increases significantly at higher $\eta$ compared with PLIV.
As observed for muons, the prompt electron efficiency decreases only mildly with increasing pile-up, while the rejection remains approximately constant, with a very slight rise at higher detector occupancy for the PLIT OPs defined to match the corresponding PLIV prompt electron efficiency.

\begin{figure}[h!]
\centering
\includegraphics[width=0.49\textwidth]{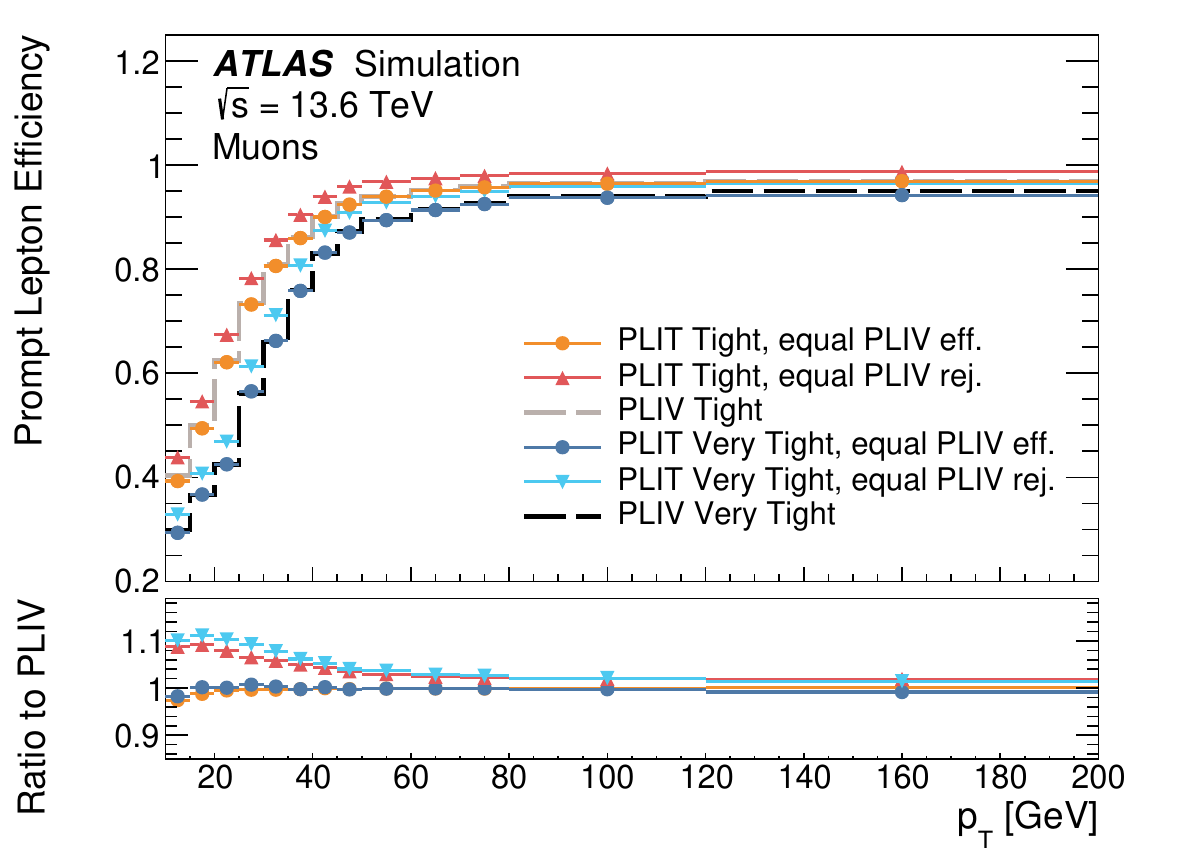}
\includegraphics[width=0.49\textwidth]{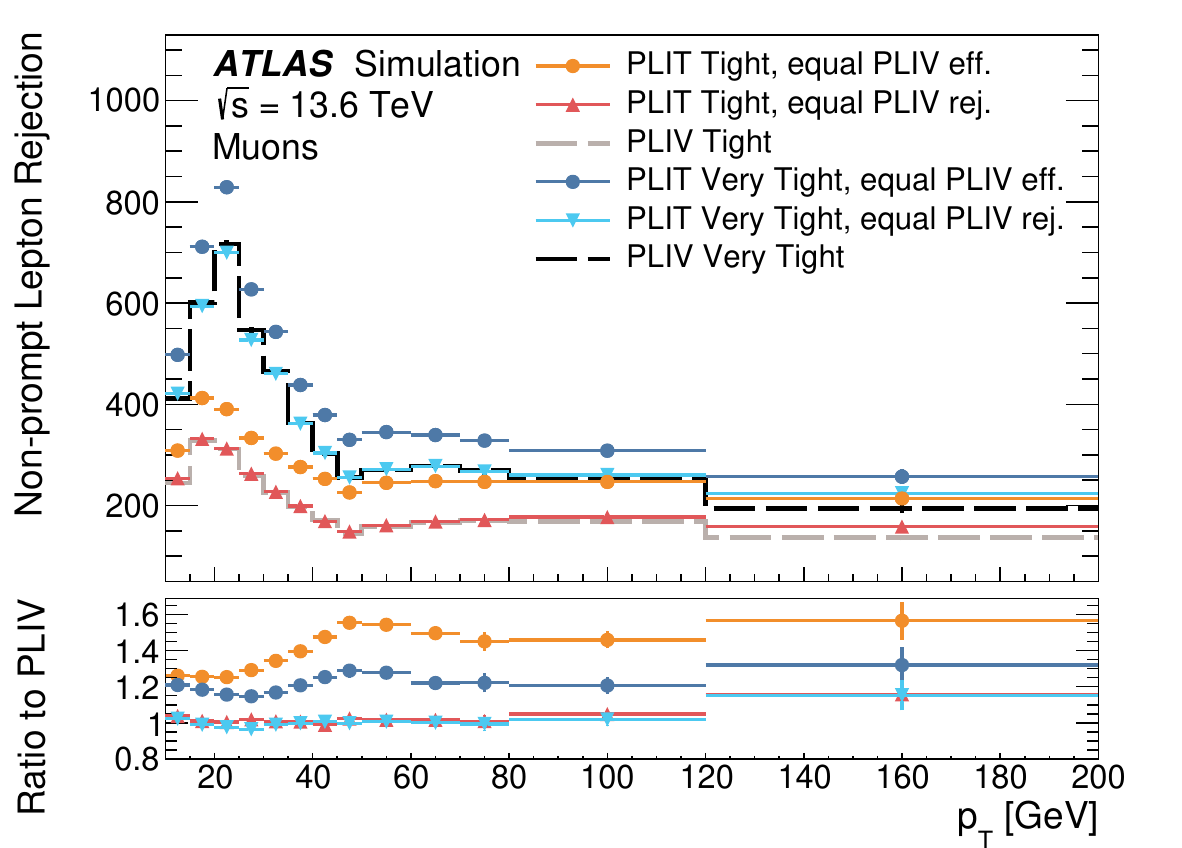}
\includegraphics[width=0.49\textwidth]{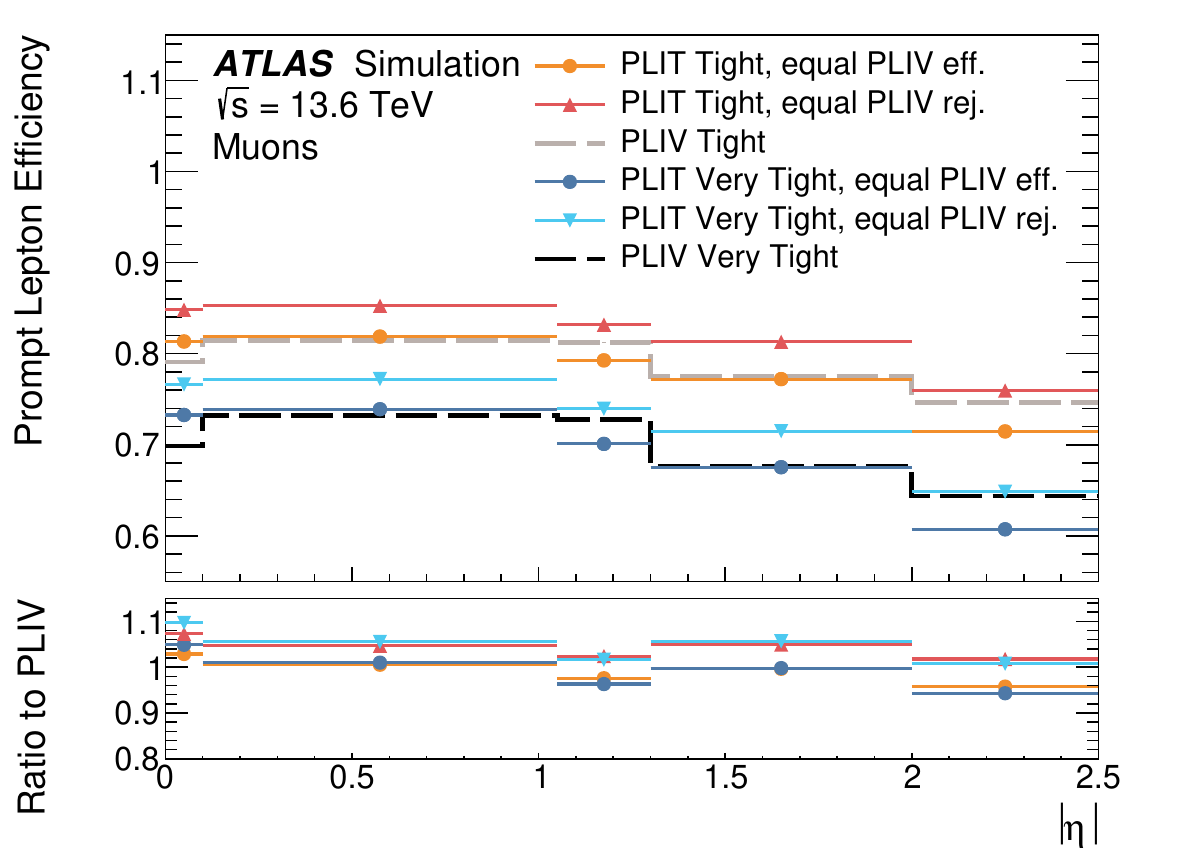}
\includegraphics[width=0.49\textwidth]{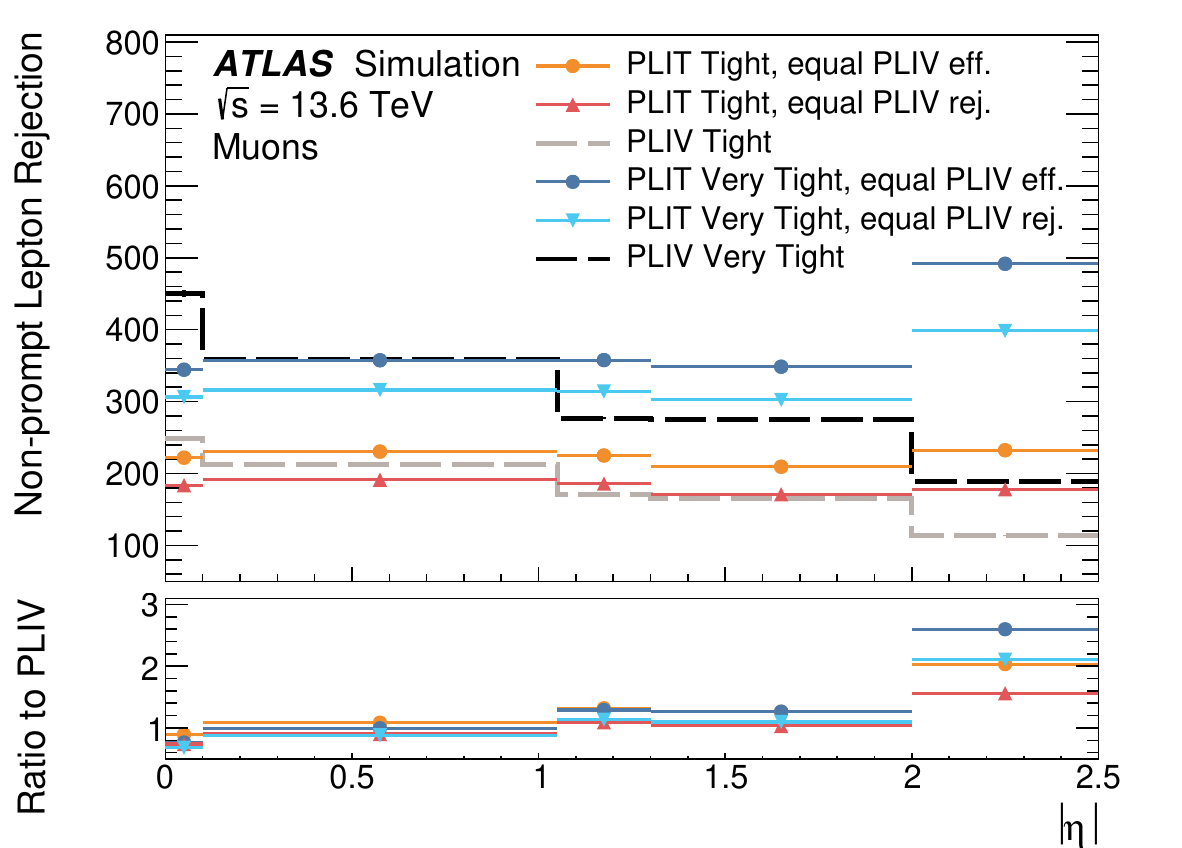}
\includegraphics[width=0.49\textwidth]{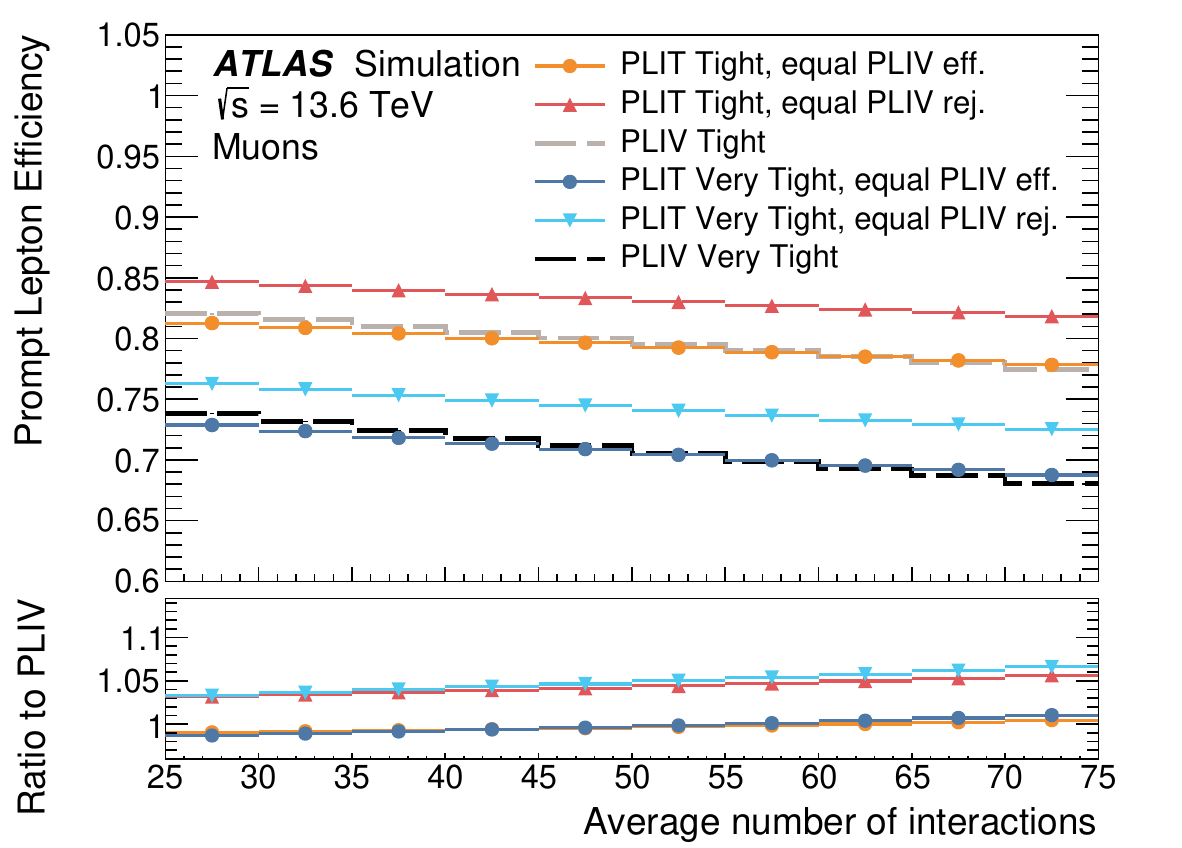}
\includegraphics[width=0.49\textwidth]{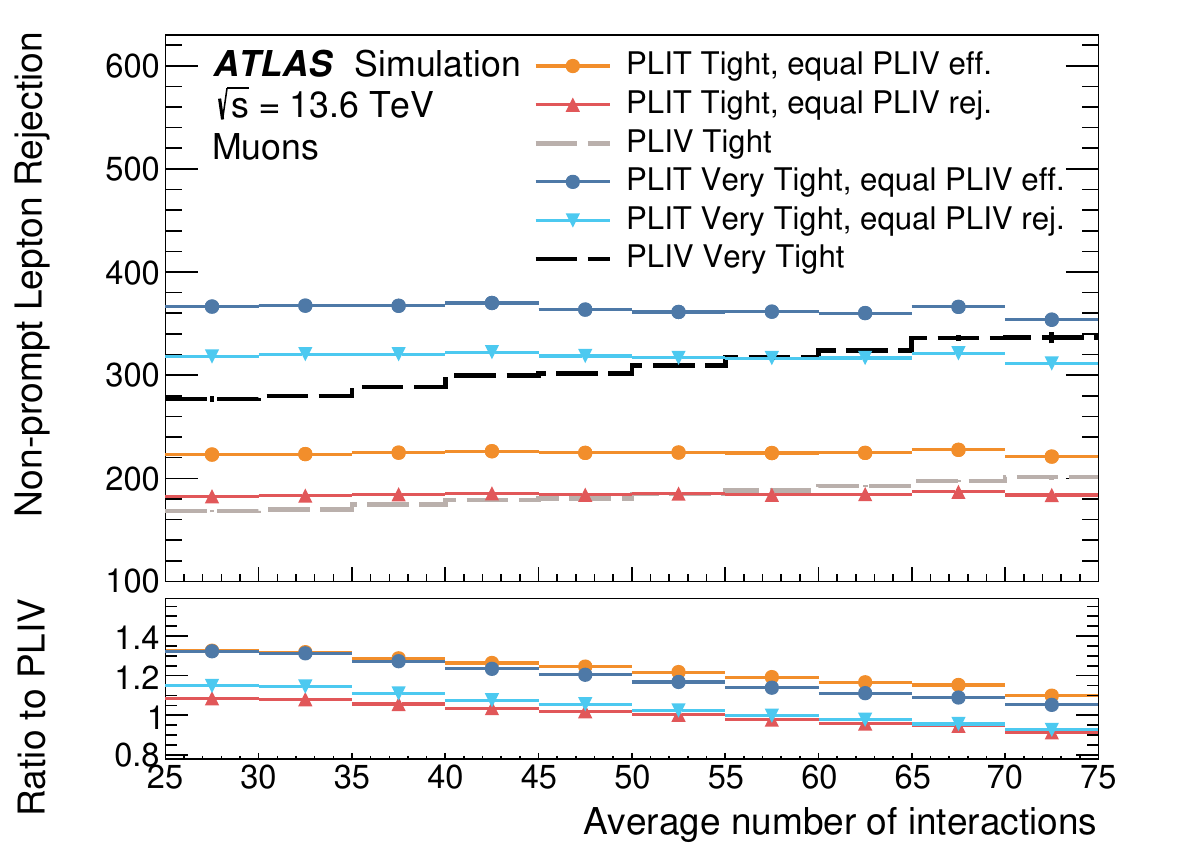}
\caption{Prompt muon efficiency (left) and non-prompt muon rejection (right)
in simulated $\ttbar$ events
for the OPs of PLIT and PLIV, evaluated for muons satisfying the
\textsc{Medium} identification OP, as a function of muon $\pt$ (top), muon
$|\eta|$ (middle), and average number of interactions per bunch crossing
(bottom).}
\label{fig:wp_mu}
\end{figure}

\begin{figure}[h!]
\centering
\includegraphics[width=0.49\textwidth]{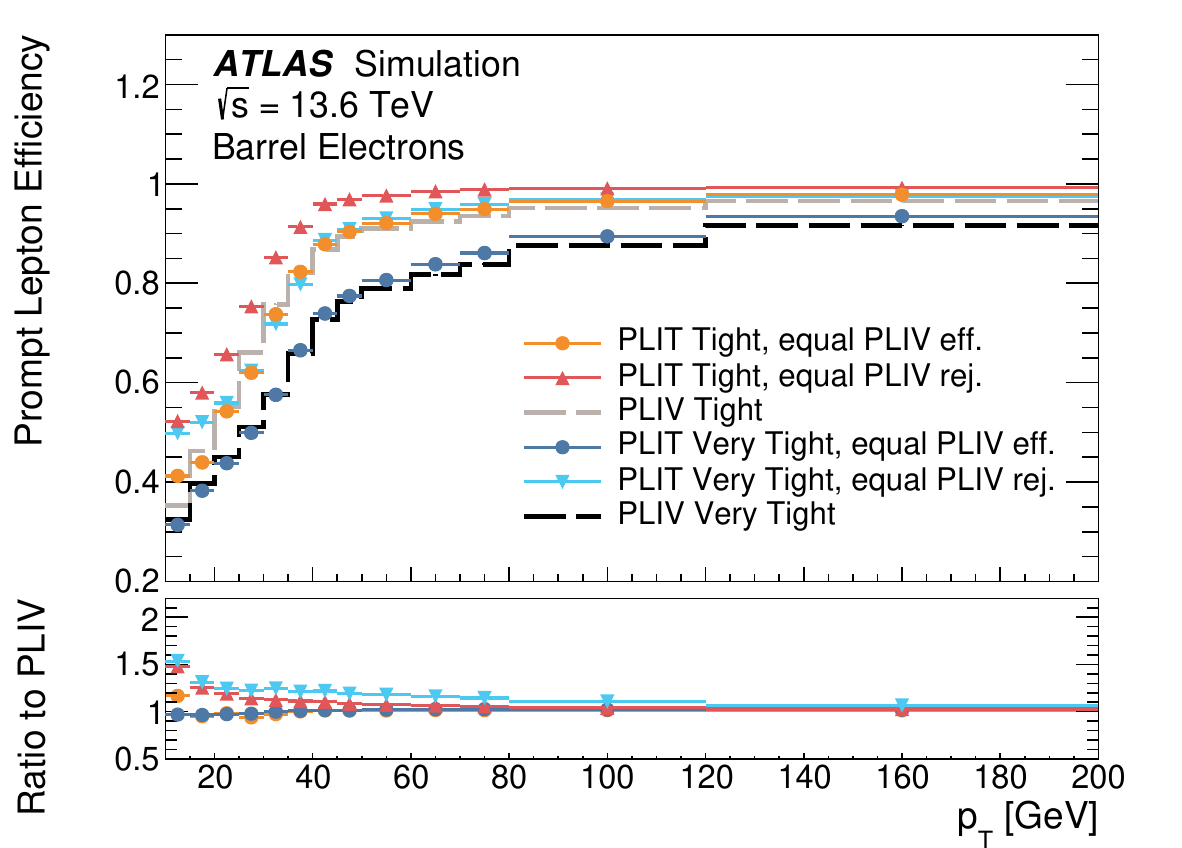}
\includegraphics[width=0.49\textwidth]{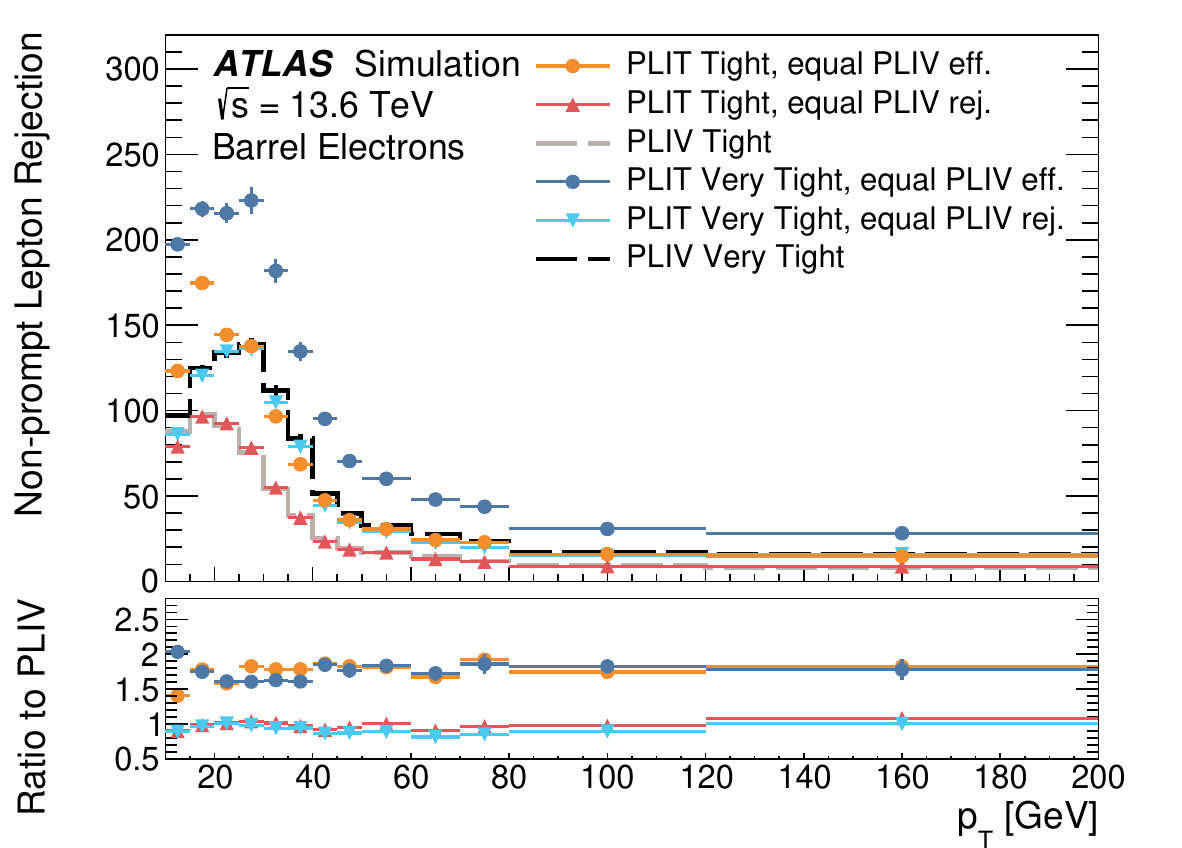}
\includegraphics[width=0.49\textwidth]{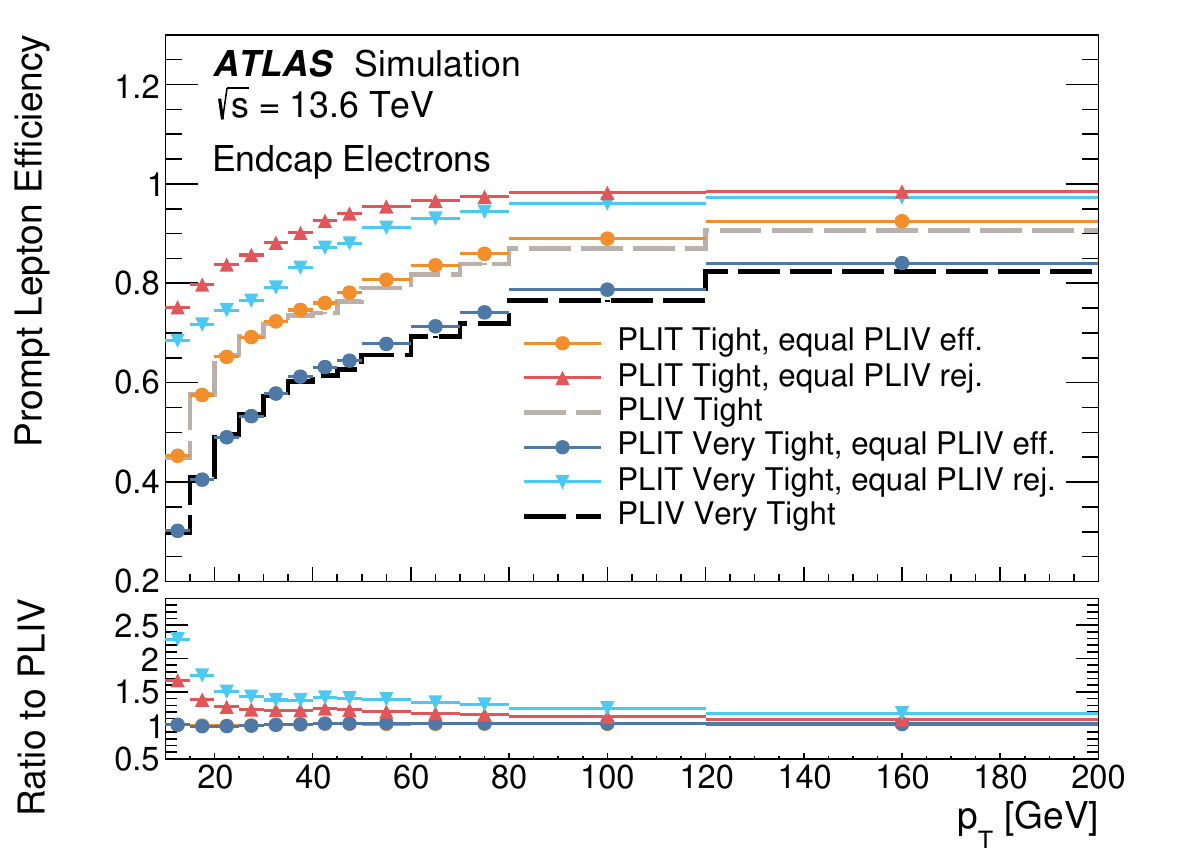}
\includegraphics[width=0.49\textwidth]{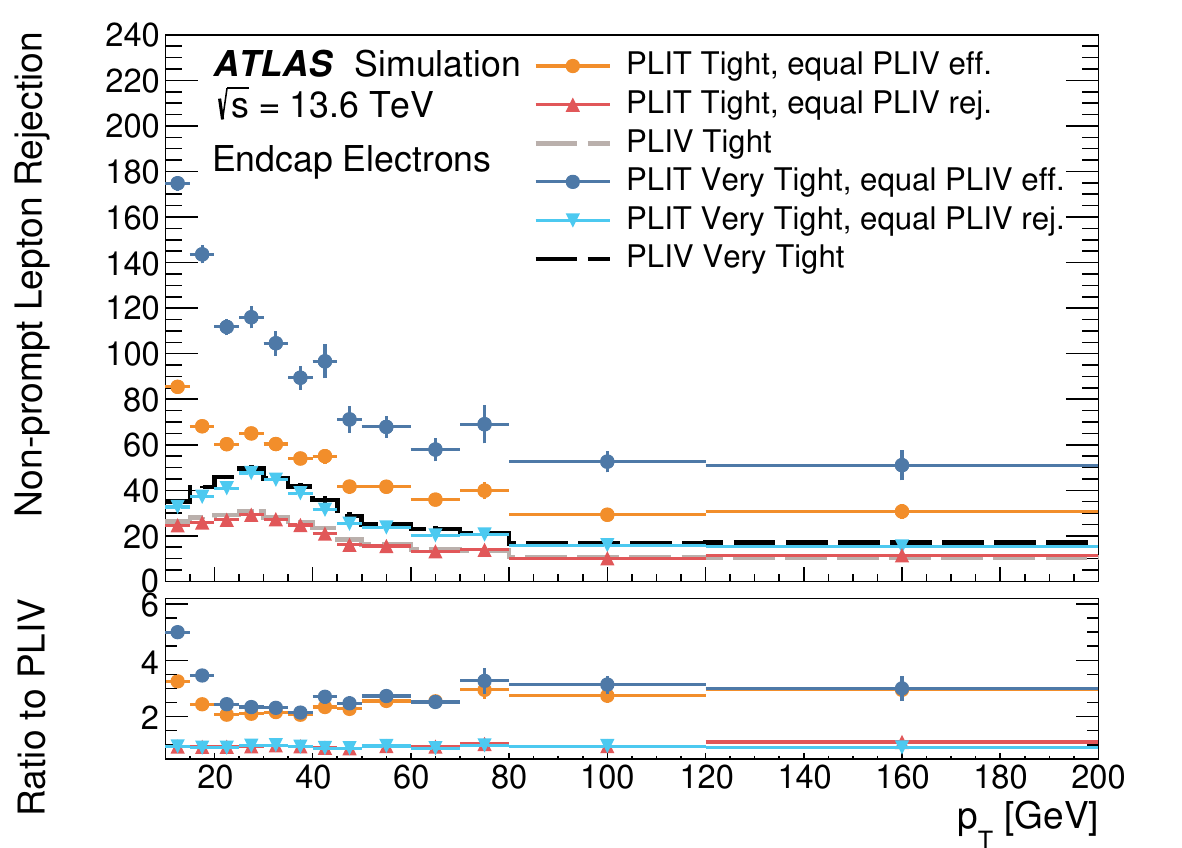}
\caption{Prompt electron efficiency (left) and non-prompt electron rejection
(right) in simulated $\ttbar$ events for the OPs of PLIT and PLIV, evaluated for electrons satisfying the
\textsc{Medium} identification OP, as a function of electron $\pt$ for barrel
electrons (top) and endcap electrons (bottom).}
\label{fig:wp_elec_pt}
\end{figure}

\begin{figure}[h!]
\centering
\includegraphics[width=0.49\textwidth]{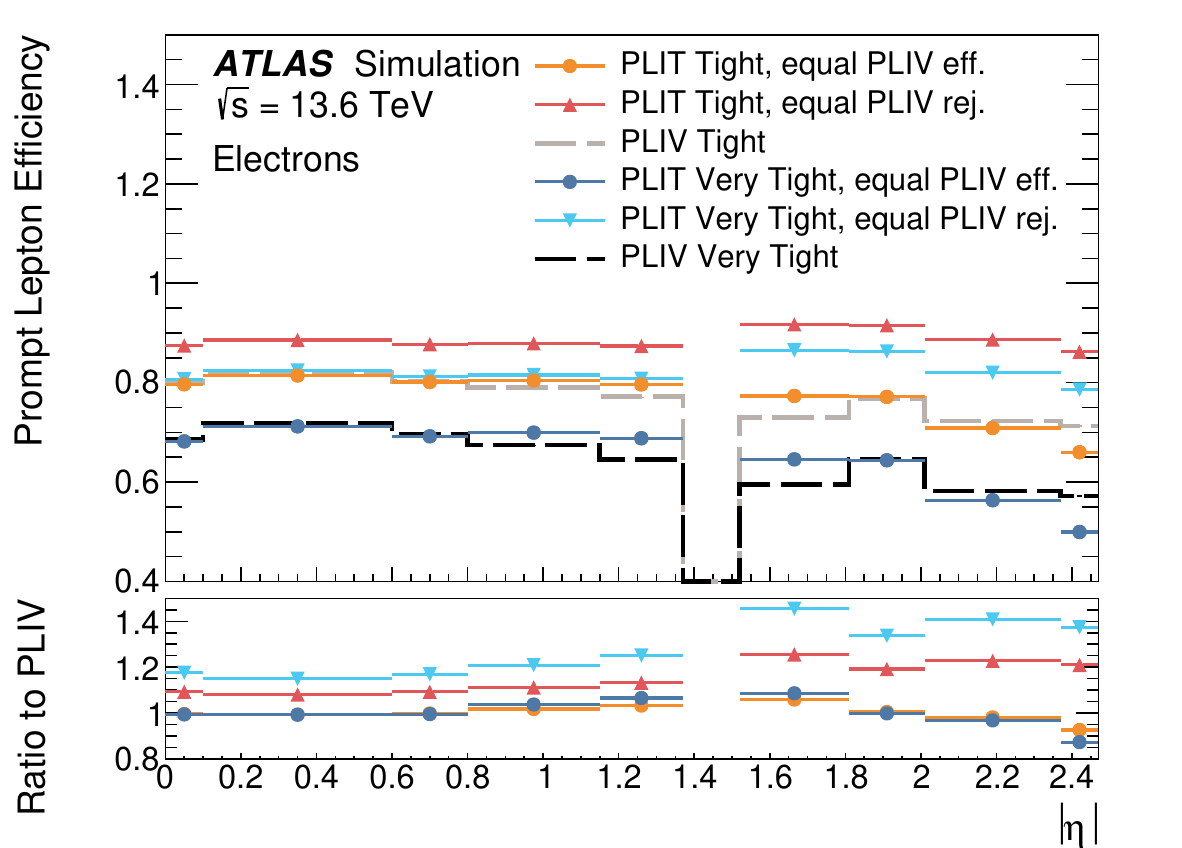}
\includegraphics[width=0.49\textwidth]{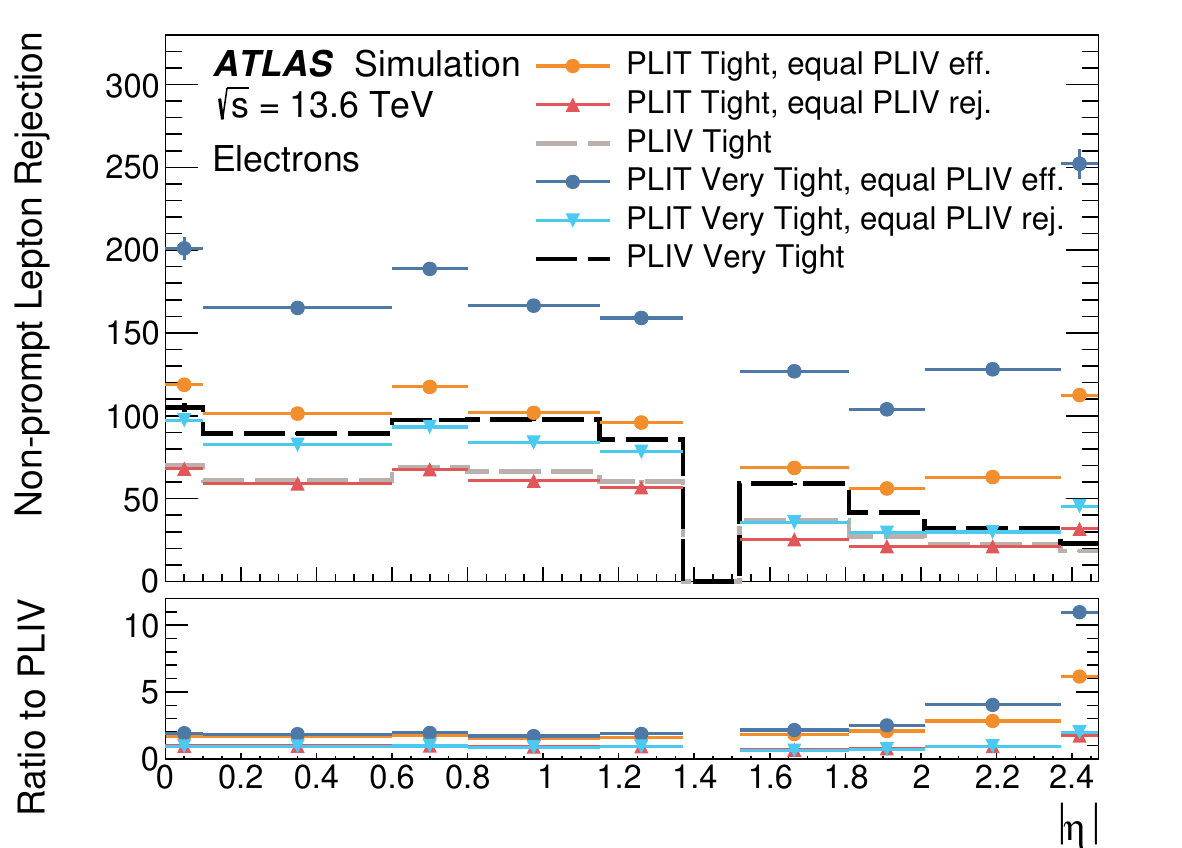}
\includegraphics[width=0.49\textwidth]{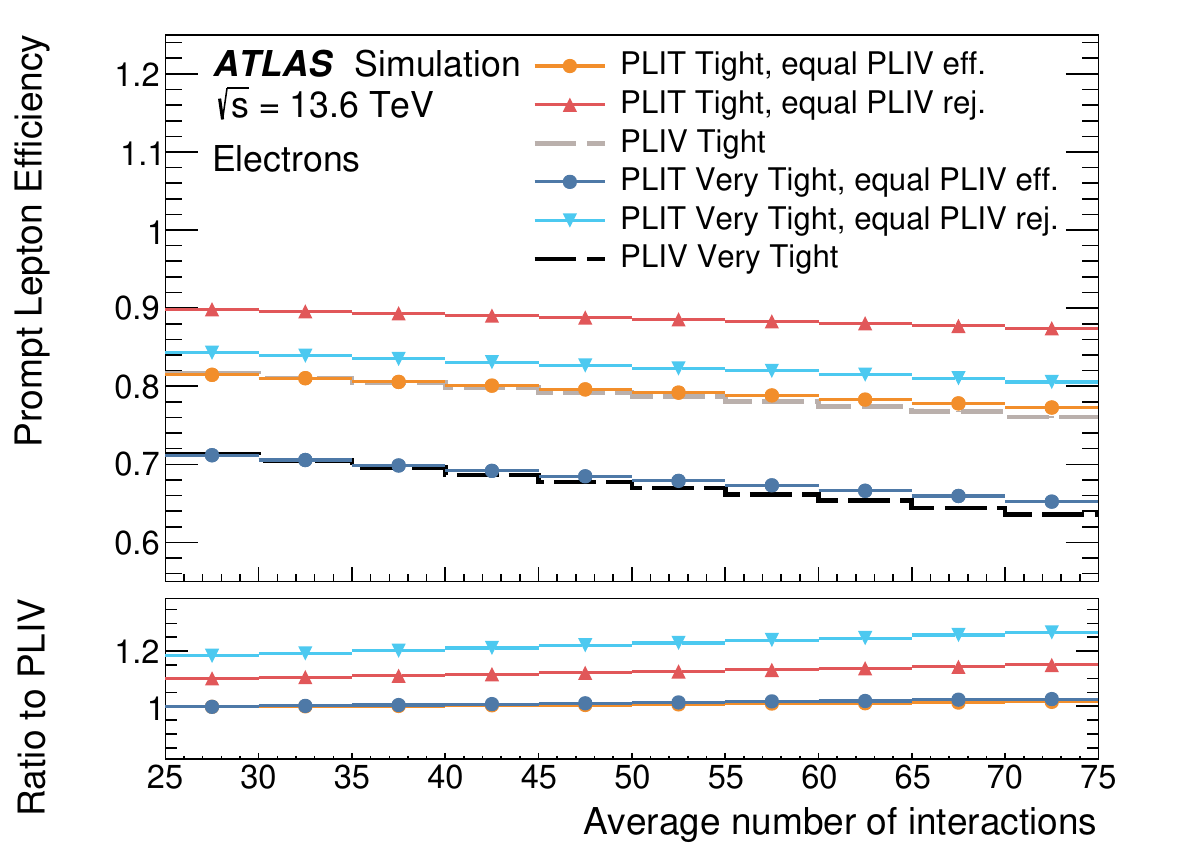}
\includegraphics[width=0.49\textwidth]{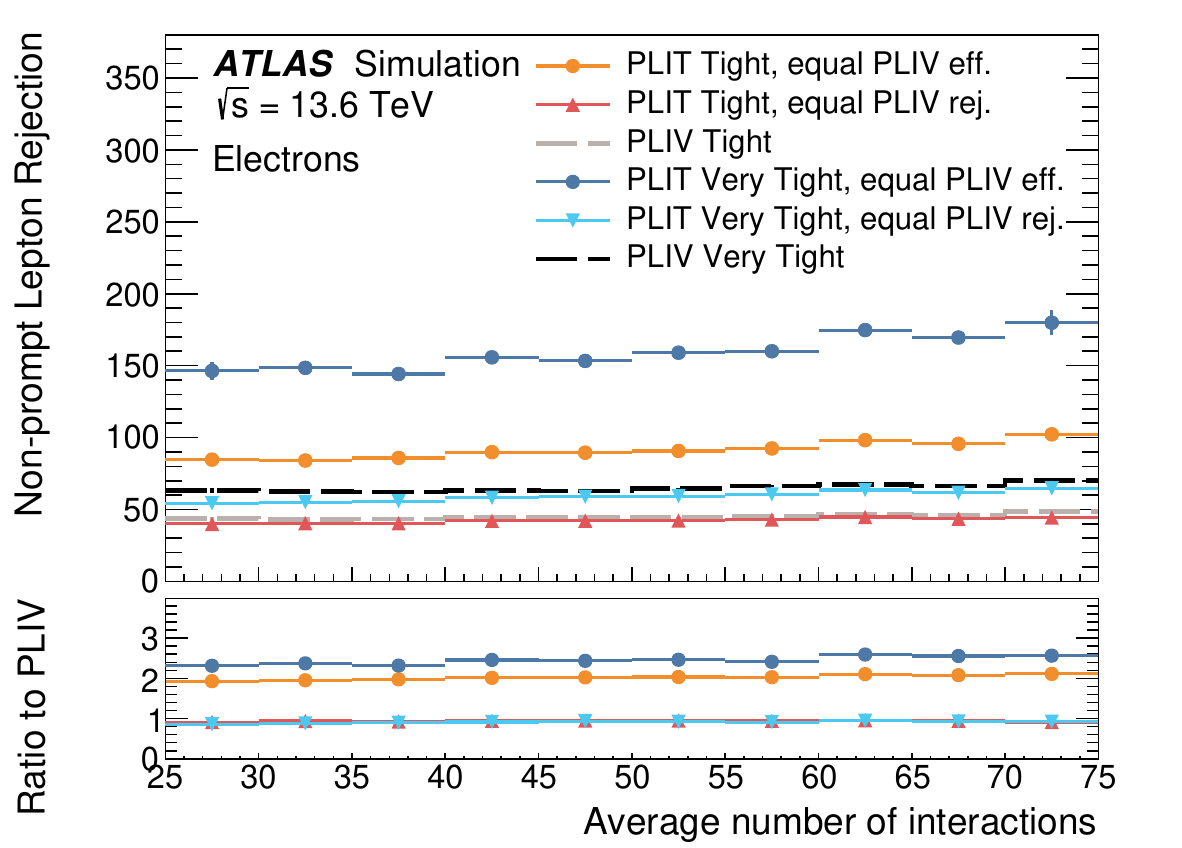}
\caption{Prompt electron efficiency (left) and non-prompt electron rejection
(right) in simulated $\ttbar$ events for the OPs of PLIT and PLIV, evaluated for electrons satisfying the
\textsc{Medium} identification OP, as a function of electron $|\eta|$ (top) and
average number of interactions per bunch crossing (bottom).}
\label{fig:wp_elec_eta_mu}
\end{figure}


%
\section{Data-to-simulation correction factors}
\label{sec:calibration}
%

The PLIT prompt lepton efficiency for each OP is calibrated in data using the standard tag-and-probe
method applied to $\Zmm$ and $\Zee$ events~\cite{MUON-2018-03,EGAM-2021-01}.
For illustration, only the calibration for the equal PLIV efficiency OPs is presented in this paper;
other calibrations will also be available to ATLAS physics analyses.
In this method, a well-identified
and triggered lepton (the \textit{tag}) is paired with a second lepton (the
\textit{probe}) originating from the same $Z$ boson decay.
Since the probe selection does not rely on the isolation requirement under study, it provides
an unbiased sample for measuring the efficiency of the PLIT OPs
in data and simulation.
A fit to the lepton pair reconstructed mass allows to extract the prompt lepton signal and subtract the non-prompt lepton component produced by the QCD multijet background,
whose mass spectrum is estimated from data using dedicated control regions.
The ratio of the efficiency in data to that in simulation defines the scale factor (SF),
which can be later applied to simulated samples.

The probe leptons are required to satisfy the requirements described in Section~\ref{sec:detector} in terms of $\pt$, pseudorapidity,
and impact parameters. In the case of electrons, probe electrons found to be within $0.2<\Delta R<0.4$ from any jet reconstructed
with the \antikt~algorithm~\cite{Cacciari:2008gp, Fastjet} using a radius parameter of 0.4 with $\pt>20$~\GeV are discarded.
The calibrations shown here are for the \textsc{Medium} identification OP.
Figure~\ref{fig:physics:wp_comp_run3} presents the muon efficiencies and SFs of the Tight
and VeryTight OPs, defined at equal PLIV efficiency, in data and
simulation for each Run~3 data-taking year.
Figure~\ref{fig:physics:wp_comp_run3_electron} shows the corresponding
efficiencies and SFs for electrons.
Since the jet environment in $\Zll$ events differs significantly from that in
$\ttbar$ events, the measured efficiencies are slightly larger than those
obtained in the $\ttbar$ samples used to present the working points in the previous section.
It is confirmed that the SFs derived in $\Zll$ are nevertheless applicable to events with more jet activity by comparing the PLIT score in data and MC
in a $\ttbar$ dilepton sample with high prompt lepton purity.
The efficiencies decrease by a few \% over time, which is attributed to the increase in pile-up.
For both muons and electrons, the SFs are measured in the range
$10 < \pt < 150$~\GeV and remain close to unity throughout the measured
$\pt$ range. The largest deviations from unity are observed at low
$\pt$, where the associated uncertainties also increase.
For both muons and electrons, SFs have residual differences of a few \% across years,
so separate SFs are provided to physics analyses for each data-taking year.
The muon SFs show no significant variation with pseudorapidity and are
therefore parameterised only as a function of $\pt$.
The electron SFs provided to physics analyses are parameterised in two
dimensions in bins of $\eta$ and $\pt$, excluding the calorimeter
transition region $1.37 < |\eta| < 1.52$.

\begin{figure}
\centering
\includegraphics[width=0.49\textwidth]{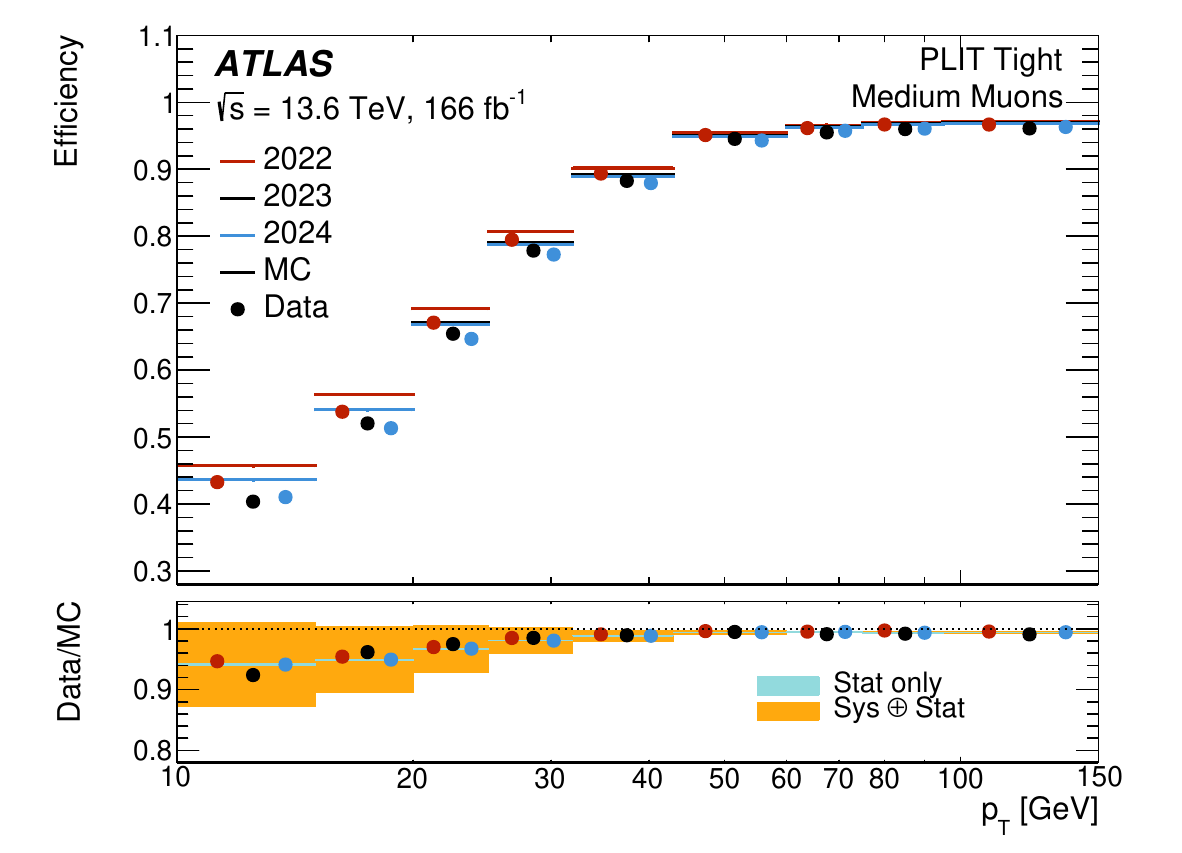}
\includegraphics[width=0.49\textwidth]{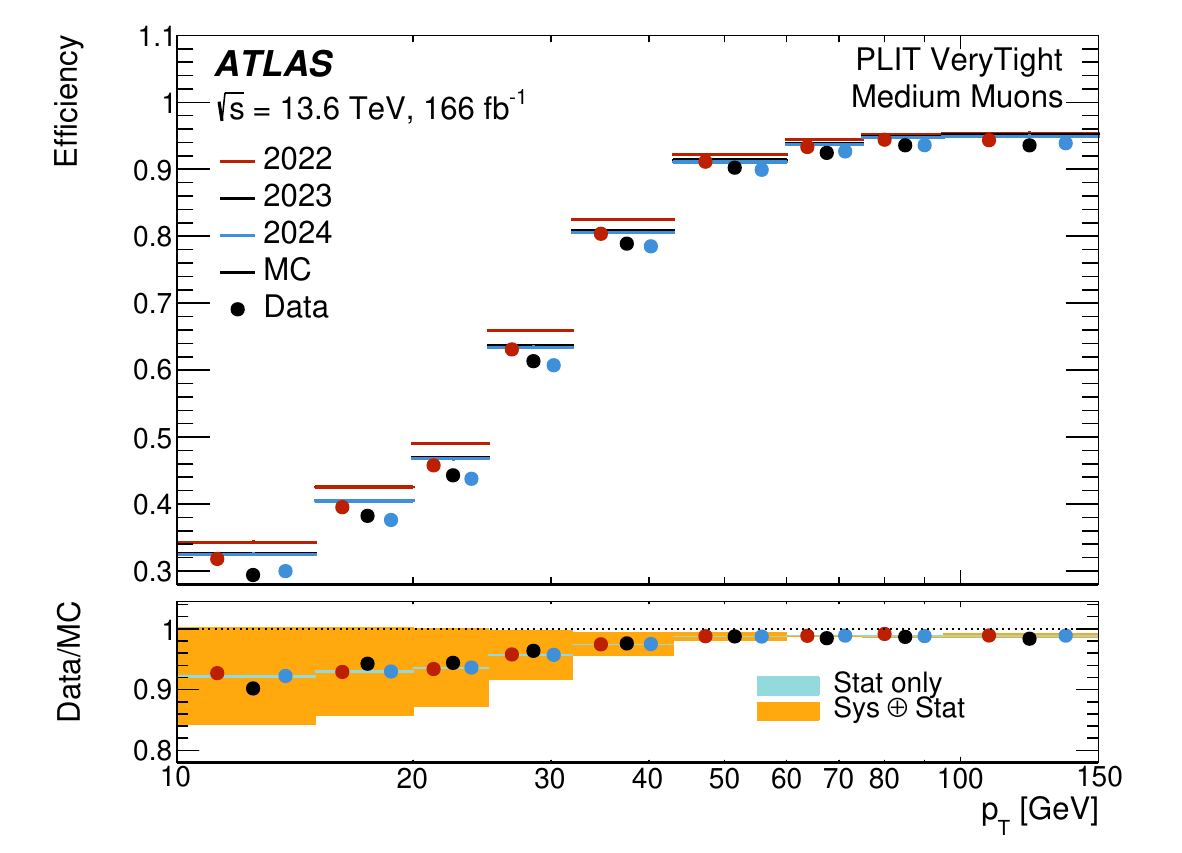}
\caption{
Data-simulation comparison of PLIT prompt muon efficiency and data-to-MC SFs
as a function of $\pt$ for muons satisfying the \textsc{Medium} identification OP,
for the Tight (left) and VeryTight (right) PLIT OPs at equal PLIV efficiency for each
data-taking year. The blue and yellow bands indicate the statistical and
total uncertainties in the SFs for the year 2024 (uncertainties for the
other years are very similar).
}
\label{fig:physics:wp_comp_run3}
\end{figure}

\begin{figure}
\centering
\includegraphics[width=0.49\textwidth]{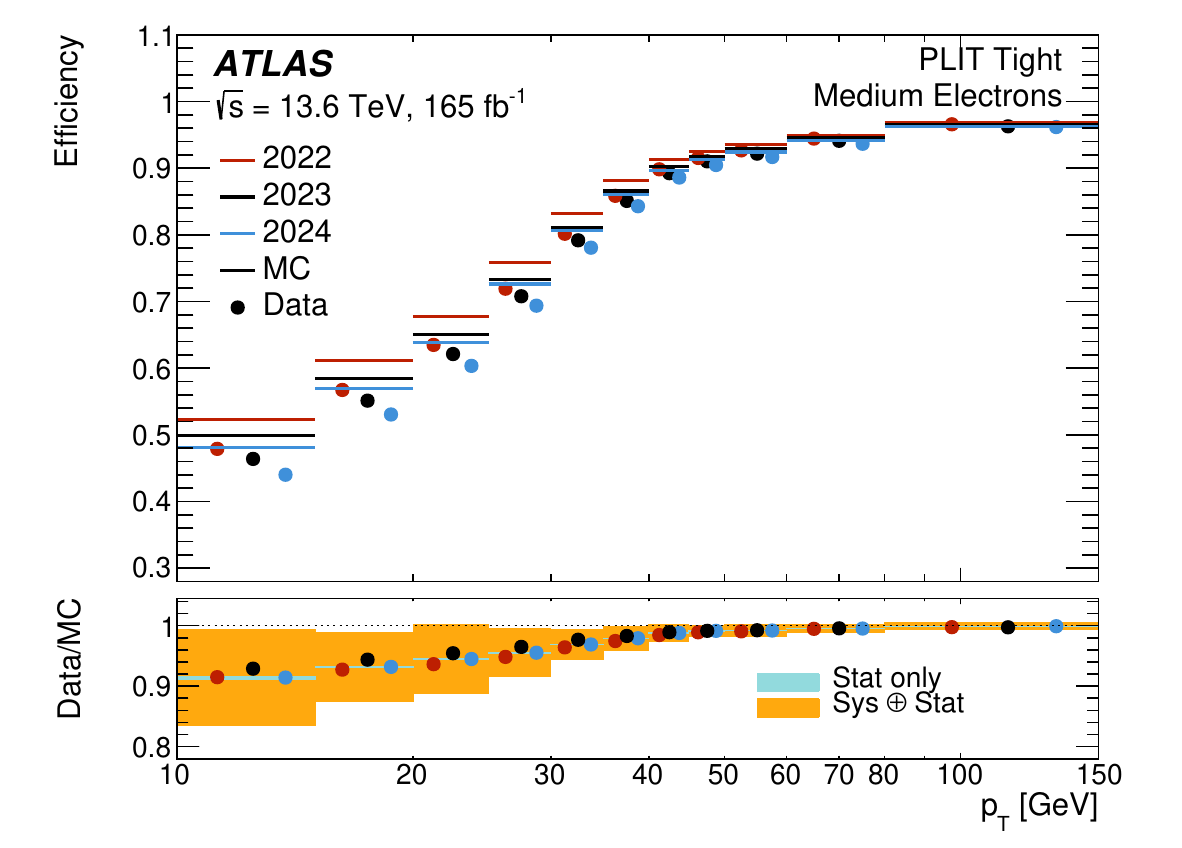}
\includegraphics[width=0.49\textwidth]{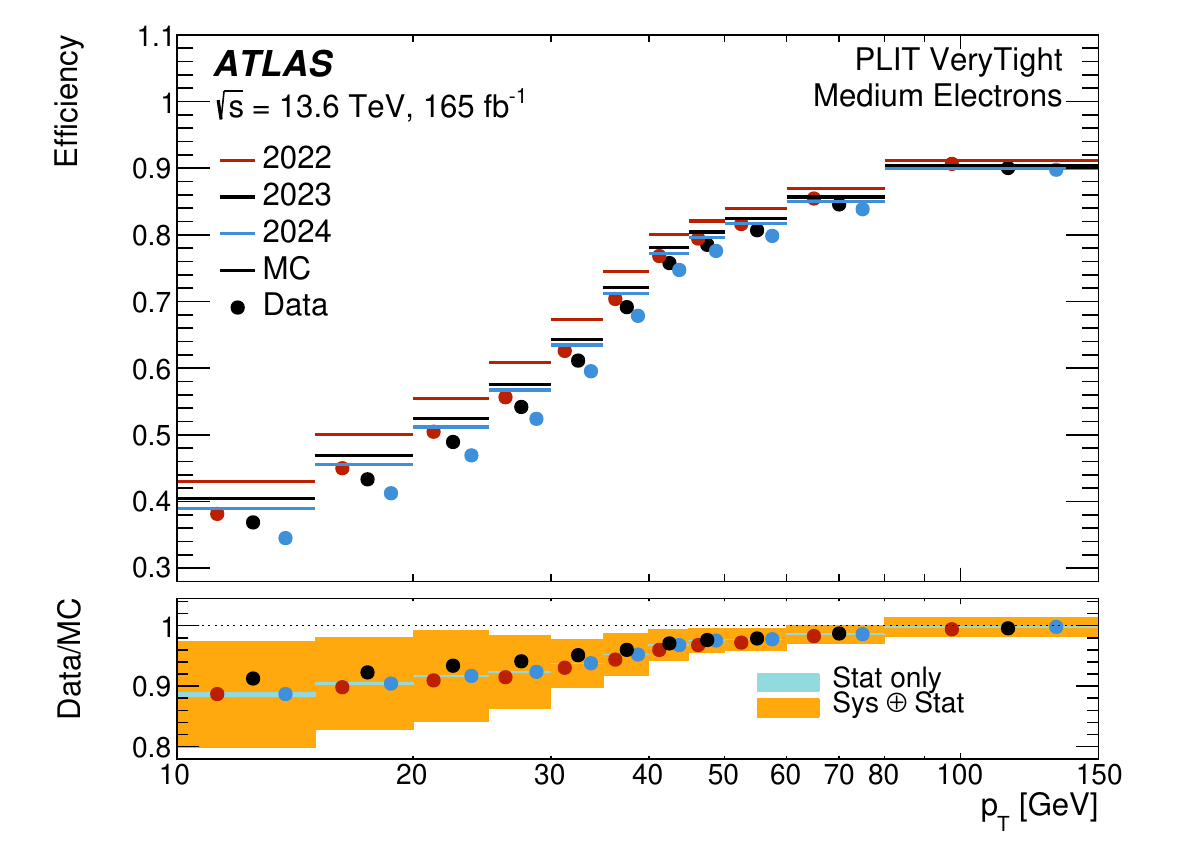}
\caption{
Data-simulation comparison of PLIT prompt electron efficiency and data-to-MC SFs
as a function of $\pt$ for electrons satisfying the \textsc{Medium} identification OP,
for the Tight (left) and VeryTight (right) PLIT OPs at equal PLIV efficiency for each
data-taking year. The blue and yellow bands indicate the statistical and
total uncertainties in the SFs for the year 2024 (uncertainties for the
other years are very similar).
}
\label{fig:physics:wp_comp_run3_electron}
\end{figure}

The systematic uncertainties are evaluated following the methodologies adopted
in the previous muon and electron isolation calibrations~\cite{MUON-2018-03,EGAM-2021-01}. Figure~\ref{fig:physics:wp_sys_pt_eff}
shows the individual uncertainty contributions for the 2024 calibration of the muon and electron VeryTight OPs
derived at efficiency equal to PLIV efficiency.
The uncertainty composition is very similar for all data-taking
years and for the different OPs, and is also consistent with what was observed for PLIV.
The dominant uncertainty for
both muons and electrons originates from the modelling of hadronic activity in
the simulation (\textit{MC modelling}). It is derived from the
difference between the nominal Powheg+Pythia simulation and an alternative
Sherpa sample. This uncertainty dominates at all values of $\pt$, decreasing
from about 10\% at $\pt\simeq10$~\GeV to 2\% above 40~\GeV. A second common source of uncertainty is associated with the choice of the
invariant-mass window used in the tag-and-probe measurement. For muons, the
nominal $81<m_{\mu\mu}<101$~\GeV window is varied to $86$--$96$~\GeV and
$76$--$106$~\GeV, resulting in the second largest uncertainty contribution,
reaching 0.3\% at low $\pt$. For electrons, the nominal $75<m_{ee}<105$~\GeV window is varied to
$70$--$110$~\GeV and $80$--$100$~\GeV, leading to a similarly small
contribution. The statistical uncertainty from the limited size of the available
data and simulated samples (\textit{Statistics})
contributes at most 0.2\% and 0.3\% respectively for muons and electrons at 10~\GeV. It is
largest at low $\pt$, reaches a minimum at intermediate $\pt$, and increases
again at the highest $\pt$ values, owing to the $\pt$ spectrum of leptons coming from $Z$ decay, which peaks around 45~\GeV, half the mass
of the $Z$ boson.

For muons, all remaining systematic uncertainties are sub-dominant. These include uncertainties
related to the signal extraction, namely the template shape
(\textit{Template shape}), the description of the QCD multijet background (\textit{QCD fit}), and the choice of the shape parameter used in the QCD fit
(\textit{$\Lambda$-SC}), together with the normalisation of simulated
backgrounds arising from the integrated luminosity
(\textit{Luminosity}) and the theoretical cross-sections of the simulated
processes (\textit{$\sigma$(MC)}). All of these uncertainties remain below 0.1\% over the full $\pt$ range.

For electrons, the second largest group of uncertainties originates from the
background modelling and comprises variations of the background control region
(\textit{Bkg shape}), the QCD multijet background (\textit{QCD fit}), and the fit range (\textit{Fit range}). Among these, the uncertainty
associated with the fit range is the dominant background-modelling contribution
up to $\pt \approx 30$~\GeV, while the remaining background-modelling
uncertainties stay below 0.1\% over the full kinematic range. Additional
signal-modelling uncertainties, arising from residual signal contamination
(\textit{Signal contamination}) and the modelling of the $Z$ boson mass
distribution (\textit{Z lineshape}), are also sub-dominant and remain below
0.1\% across the full measured phase space.

\begin{figure}
\centering
\includegraphics[width=0.49\textwidth]{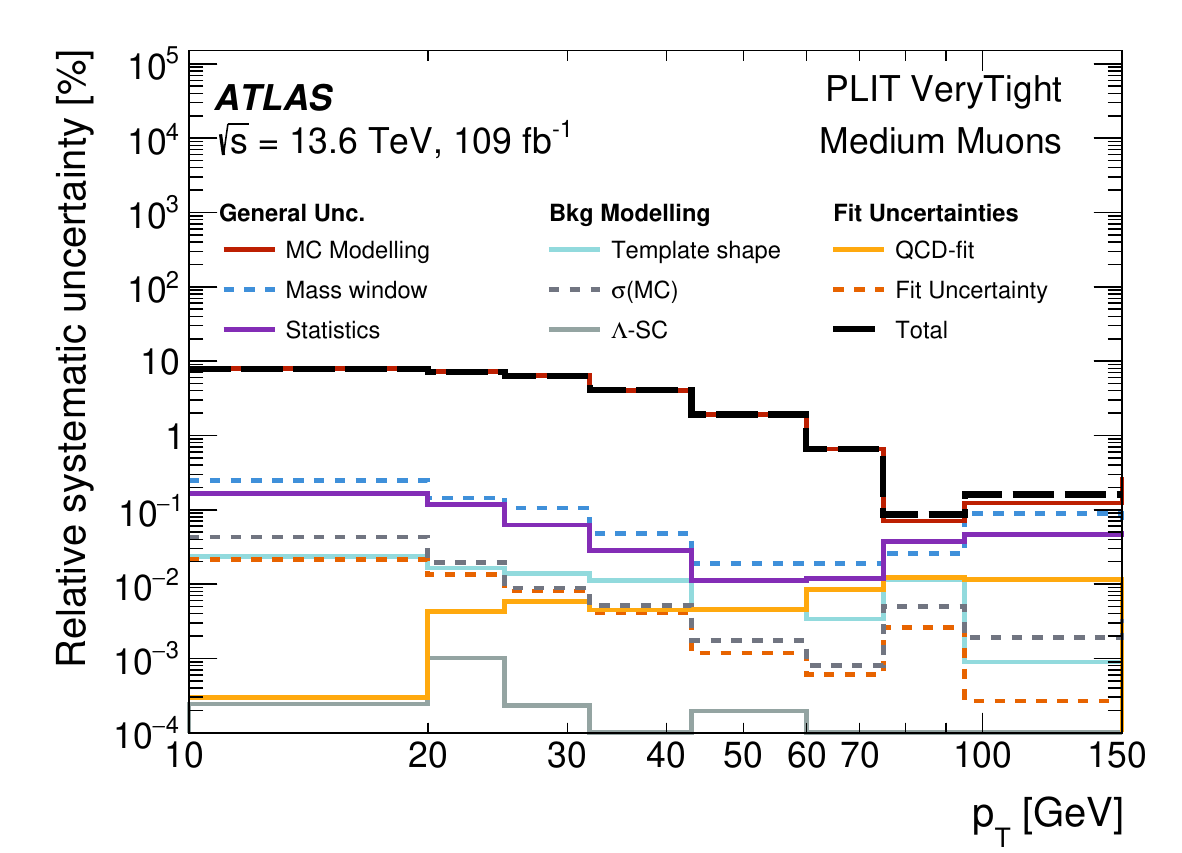}
\includegraphics[width=0.49\textwidth]{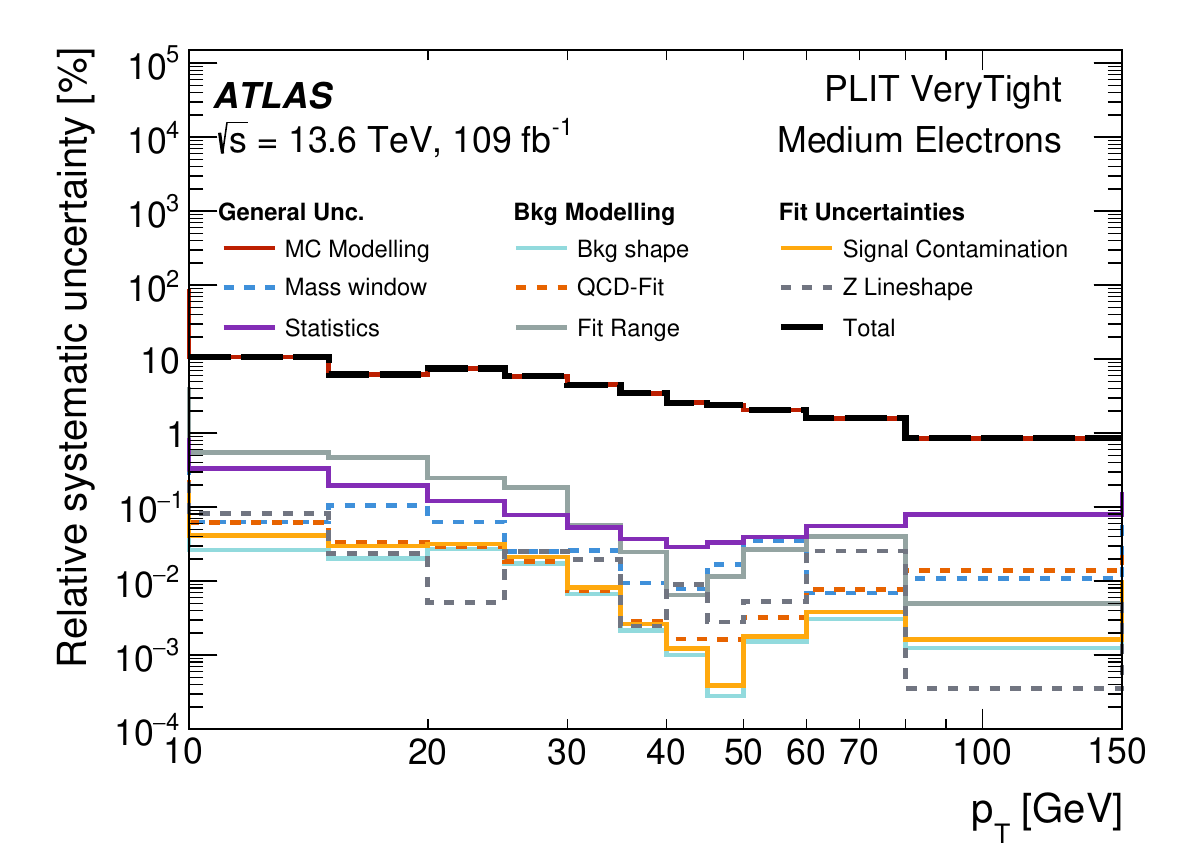}
\caption{
Contributions of systematic uncertainties per $\pt$ bin for the muon (left)
and electron (right) VeryTight OPs, using leptons satisfying the
\textsc{Medium} identification OP, derived at efficiency equal to PLIV
efficiency for the 2024 data-taking year of the ATLAS Run~3 campaign.
}
\label{fig:physics:wp_sys_pt_eff}
\end{figure}


%
%
\FloatBarrier

\section{Conclusion}
\label{sec:conclusion}
%

A transformer-based neural network algorithm has been developed to distinguish
prompt leptons (muons and electrons) produced in electroweak boson decays,
from non-prompt leptons produced predominantly in heavy-flavour hadron decays.
The Prompt Lepton Isolation Tagger adopts the architecture of the ATLAS
jet flavour-tagging algorithm and combines conventional isolation observables
with detailed track-level information sensitive to displaced heavy-flavour
decays. The transformer architecture, together with auxiliary track-origin
classification and vertexing tasks, enables this information to be exploited
effectively, leading to significantly improved discrimination compared with
both the previous BDT-based multivariate isolation algorithm and the conventional
calorimeter- and track-based isolation criteria.
Operating points have been defined and their prompt lepton efficiencies have been calibrated in data using
the standard $\Zll$ tag-and-probe method.
For the same non-prompt lepton rejection, a relative improvement of the prompt lepton efficiency of 10\% for muons, 50\% for barrel electrons, and 100\% for endcap electrons, is achieved at low $\pt$.
A similar performance is observed for the algorithm trained on Run~2 simulation.
This significant performance improvement will benefit measurements relying on multilepton signatures.


%
%
%
%
%
\section*{Acknowledgements}
%

%

%
%

%
%

We thank CERN for the very successful operation of the LHC and its injectors, as well as the support staff at
CERN and at our institutions worldwide without whom ATLAS could not be operated efficiently.

The crucial computing support from all WLCG partners is acknowledged gratefully, in particular from CERN, the ATLAS Tier-1 facilities at TRIUMF/SFU (Canada), NDGF (Denmark, Norway, Sweden), CC-IN2P3 (France), KIT/GridKA (Germany), INFN-CNAF (Italy), NL-T1 (Netherlands), PIC (Spain), RAL (UK) and BNL (USA), the Tier-2 facilities worldwide and large non-WLCG resource providers. Major contributors of computing resources are listed in Ref.~\cite{ATL-SOFT-PUB-2026-001}.

We gratefully acknowledge the support of ANPCyT, Argentina; YerPhI, Armenia; ARC, Australia; BMWFW and FWF, Austria; ANAS, Azerbaijan; CNPq and FAPESP, Brazil; NSERC, NRC and CFI, Canada; CERN; ANID, Chile; CAS, MOST, NSFC and FRFCU, China; Minciencias, Colombia; MEYS CR, Czech Republic; DNRF and DNSRC, Denmark; IN2P3-CNRS and CEA-DRF/IRFU, France; SRNSFG, Georgia; BMFTR, HGF and MPG, Germany; GSRI, Greece; RGC and Hong Kong SAR, China; ICHEP and Academy of Sciences and Humanities, Israel; INFN, Italy; MEXT and JSPS, Japan; CNRST, Morocco; NWO, Netherlands; RCN, Norway; MNiSW, Poland; FCT, Portugal; MNE/IFA, Romania; MSTDI, Serbia; MSSR, Slovakia; ARIS and MVZI, Slovenia; DSI/NRF, South Africa; MICIU/AEI, Spain; SRC and Wallenberg Foundation, Sweden; SERI, SNSF and Cantons of Bern and Geneva, Switzerland; NSTC, Taipei; TENMAK, T\"urkiye; STFC/UKRI, United Kingdom; DOE and NSF, United States of America.

Individual groups and members have received support from BCKDF, CANARIE, CRC and DRAC, Canada; CERN-CZ and PRIMUS, Czech Republic; COST, ERC, ERDF, Horizon 2020 and Marie Sk{\l}odowska-Curie Actions, European Union; Investissements d'Avenir Labex, Investissements d'Avenir Idex and ANR, France; DFG and AvH Foundation, Germany; Herakleitos, Thales and Aristeia programmes co-financed by EU-ESF and the Greek NSRF, Greece; BSF-NSF and MINERVA, Israel; NCN and NAWA, Poland; La Caixa Banking Foundation, CERCA and AGAUR programs from Generalitat de Catalunya and PROMETEO and GenT Programmes Generalitat Valenciana, Spain; G\"{o}ran Gustafssons Stiftelse, Sweden; The Royal Society and Leverhulme Trust, United Kingdom; Eric and Wendy Schmidt Fund for Strategic Innovation, United States of America.

In addition, individual members wish to acknowledge support from Chile: Agencia Nacional de Investigaci\'on y Desarrollo (ANID FONDECYT reg. 1230987, FONDECYT 1230812, FONDECYT 1240864, Fondecyt 3240661, Fondecyt Regular 1240721); China: Fundamental Research Funds for the Central Universities (010-63263105), Chinese Ministry of Science and Technology (MOST-2023YFA1605700, MOST-2023YFA1609300), National Natural Science Foundation of China (NSFC 12275265, NSFC-W2543005); Czech Republic: Czech Science Foundation (GACR - 24-11373S), Ministry of Education Youth and Sports (ERC-CZ-LL2327, FORTE CZ.02.01.01/00/22\_008/0004632); EU: H2020 European Research Council (ERC - 101002463); European Union: European Research Council (BARD No. 101116429, ERC - 101219398, ERC 101089007), European Regional Development Fund (HE COFUND GA No.101081355, ERDF), Marie Sklodowska-Curie Actions (GAP-101168829); France: Agence Nationale de la Recherche (ANR-22-EDIR-0002, ANR-24-CE31-0504-01); China: Research Grants Council (GRF); Italy: Istituto Nazionale di Fisica Nucleare (LHC-MIUR - 28003/2025), Ministero dell'Università e della Ricerca (NextGenEU I53D23000820006 M4C2.1.1, SOE2024\_0000023); Japan: Japan Society for the Promotion of Science (JSPS KAKENHI  JP25H0063, JSPS KAKENHI JP22H04944, JSPS KAKENHI JP24K23939, JSPS KAKENHI JP24KK0251, JSPS KAKENHI JP25H00650, JSPS KAKENHI JP25H01291, JSPS KAKENHI JP25K01011, JSPS KAKENHI JP25K01023, JSPS KAKENHI JP25KK0047); Poland: Polish National Science Centre (NCN OPUS 2023/51/B/ST2/02507, NCN OPUS nr 2022/47/B/ST2/03059, UMO-2019/34/E/ST2/00393, UMO-2021/42/E/ST2/00350, UMO-2022/47/O/ST2/00148, UMO-2023/49/B/ST2/04085, UMO-2023/51/B/ST2/00920, UMO-2024/53/N/ST2/00869); Spain: Agència de Gestió d'Ajuts Universitaris i de Recerca. (AGAUR - 2023 BP 00141), Ministry of Science and Innovation (RYC2019-028510-I, RYC2020-030254-I, RYC2021-031273-I, RYC2022-038164-I), Ministerio de Ciencia, Innovación y Universidades/Agencia Estatal de Investigaci\'on (PID2022-142604OB-C22); Sweden: Carl Trygger Foundation (Carl Trygger Foundation CTS 22:2312), Swedish Research Council (Swedish Research Council 2023-04654, VR 2021-03651, VR 2022-03845, VR 2022-04683, VR 2023-03403, VR 2024-05451, VR 2025-05940), Knut and Alice Wallenberg Foundation (KAW 2023.0366); United Kingdom: The Binks Trust, Royal Society (NIF-R1-231091); United States of America: U.S. Department of Energy (ECA DE-AC02-76SF00515), John Templeton Foundation (John Templeton Foundation 63206), Neubauer Family Foundation.

%
%


%
%
\printbibliography
\clearpage
\input{atlas_authlist}
%

%
%
%
%

%
%
%
%

\end{document}

%% file: atlas_authlist.tex
 
\begin{flushleft}
\hypersetup{urlcolor=black}
{\Large The ATLAS Collaboration}

\bigskip

\AtlasOrcid[0000-0002-6665-4934]{G.~Aad}$^\textrm{\scriptsize 102}$,
\AtlasOrcid[0000-0001-7616-1554]{E.~Aakvaag}$^\textrm{\scriptsize 17}$,
\AtlasOrcid[0000-0002-5888-2734]{B.~Abbott}$^\textrm{\scriptsize 121}$,
\AtlasOrcid[0000-0002-0287-5869]{S.~Abdelhameed}$^\textrm{\scriptsize 83b}$,
\AtlasOrcid[0000-0002-1002-1652]{K.~Abeling}$^\textrm{\scriptsize 54}$,
\AtlasOrcid[0000-0001-5763-2760]{N.J.~Abicht}$^\textrm{\scriptsize 48}$,
\AtlasOrcid[0000-0002-8496-9294]{S.H.~Abidi}$^\textrm{\scriptsize 30}$,
\AtlasOrcid[0009-0003-6578-220X]{M.~Aboelela}$^\textrm{\scriptsize 44}$,
\AtlasOrcid[0000-0002-9987-2292]{A.~Aboulhorma}$^\textrm{\scriptsize 36e}$,
\AtlasOrcid[0000-0001-5329-6640]{H.~Abramowicz}$^\textrm{\scriptsize 154}$
% ...
\bigskip
\\

% ...
$^{17}$Department for Physics and Technology, University of Bergen, Bergen; Norway.\\
% ...
$^{30}$Physics Department, Brookhaven National Laboratory, Upton NY; United States of America.\\
% ...
$^{36}$$^{(a)}$Facult\'e des Sciences Ain Chock, Universit\'e Hassan II de Casablanca;$^{(b)}$Facult\'{e} des Sciences, Universit\'{e} Ibn-Tofail, K\'{e}nitra;$^{(c)}$Facult\'e des Sciences Semlalia, Universit\'e Cadi Ayyad, LPHEA-Marrakech;$^{(d)}$LPMR, Facult\'e des Sciences, Universit\'e Mohamed Premier, Oujda;$^{(e)}$Facult\'e des sciences, Universit\'e Mohammed V, Rabat;$^{(f)}$Institute of Applied Physics, Mohammed VI Polytechnic University, Ben Guerir; Morocco.\\
% ...
$^{44}$Physics Department, Southern Methodist University, Dallas TX; United States of America.\\
% ...
$^{48}$Fakult\"{a}t Physik, Technische Universit{\"a}t Dortmund, Dortmund; Germany.\\
% ...
$^{54}$II. Physikalisches Institut, Georg-August-Universit\"{a}t G\"ottingen, G\"ottingen; Germany.\\
% ...
$^{83}$$^{(a)}$Khalifa University of Science and Technology, Abu Dhabi;$^{(b)}$New York University Abu Dhabi, Abu Dhabi;$^{(c)}$United Arab Emirates University, Al Ain;$^{(d)}$University of Sharjah, Sharjah; United Arab Emirates.\\
% ...
$^{102}$CPPM, Aix-Marseille Universit\'e, CNRS/IN2P3, Marseille; France.\\
% ...
$^{121}$Homer L. Dodge Department of Physics and Astronomy, University of Oklahoma, Norman OK; United States of America.\\
% ...
$^{154}$Raymond and Beverly Sackler School of Physics and Astronomy, Tel Aviv University, Tel Aviv; Israel.\\
% ...
$^{a}$ Also at Affiliated with an institute formerly covered by a cooperation agreement with CERN.\\
$^{b}$ Also at An-Najah National University, Nablus; Palestine.\\
$^{c}$ Also at Borough of Manhattan Community College, City University of New York, New York NY; United States of America.\\
$^{d}$ Also at Center for Interdisciplinary Research and Innovation (CIRI-AUTH), Thessaloniki; Greece.\\
$^{e}$ Also at Centre of Physics of the Universities of Minho and Porto (CF-UM-UP); Portugal.\\
$^{f}$ Also at CERN, Geneva; Switzerland.\\
% ...

\end{flushleft}
